\documentclass[twocolumn,twocolappendix]{aastex63}
\usepackage{amsmath}
\usepackage{color}
\usepackage{ulem}
\usepackage[varg]{txfonts}

\shorttitle{Transmission Spectra of Disintegrating Exoplanets}
\shortauthors{Okuya et al.}

\begin{document}

\title{Constraining the Bulk Composition of Disintegrating Exoplanets Using Combined Transmission Spectra from JWST and SPICA}

\correspondingauthor{Ayaka Okuya}
\email{okuya.a.aa@m.titech.ac.jp}

\author[0000-0001-6222-9423]{Ayaka Okuya}
\affiliation{Department of Earth and Planetary Sciences, Tokyo Institute of Technology, 2-12-1 Ookayama, Meguro-ku, Tokyo 152-8551, Japan}

\author[0000-0002-1886-0880]{Satoshi Okuzumi}
\affiliation{Department of Earth and Planetary Sciences, Tokyo Institute of Technology, 2-12-1 Ookayama, Meguro-ku, Tokyo 152-8551, Japan}

\author[0000-0003-3290-6758]{Kazumasa Ohno}
\affiliation{Department of Earth and Planetary Sciences, Tokyo Institute of Technology, 2-12-1 Ookayama, Meguro-ku, Tokyo 152-8551, Japan}

\author[0000-0003-3618-7535]{Teruyuki Hirano}
\affiliation{Department of Earth and Planetary Sciences, Tokyo Institute of Technology, 2-12-1 Ookayama, Meguro-ku, Tokyo 152-8551, Japan}




\begin{abstract}

Disintegrating planets are ultra-short-period exoplanets that appear to have a comet-like dust tail. They are commonly interpreted as low-mass planets whose solid surface is evaporating and whose tail is made of recondensing minerals.
Transmission spectroscopy of the dust tails could thus allow us to directly probe the elementary compositions of the planets. 
Previous work already investigated the feasibility of such observations using the JWST mid-infrared instrument. In this study, we explore if one can obtain a strong constrain on the tail composition by adding spectroscopy at longer wavelengths using SPICA mid-infrared instrument. We use a simple model for the spatial distribution of the dust tails and produce their synthetic transmission spectra assuming various dust compositions.
We find that combined infrared spectra from JWST and SPICA will allow us to diagnose various components of the dust tails. 
JWST will be able to detect silicate and carbide absorption features with a feature-to-noise ratio of $\ga3$
in the tail transmission spectrum of a disintegrating planet located within 100 pc from the Earth with a transit depth deeper than 0.5\%. SPICA can distinguish between Fe- and Mg-bearing crystalline silicates  for planets at $\la$ 100 pc with a transit depth of $\ga$ 2\%. Transit searches with current and future space telescopes (e.g., {\it TESS} and PLATO) will provide ideal targets for such spectroscopic observations.

\end{abstract}





\section{Introduction}\label{sec:intro}
Constraining the bulk compositions of planets is crucial for understanding their formation and evolution history.
The most common approach to infer the bulk composition of an exoplanet is to use its bulk density  \citep[e.g.,][]{Fortney+2007,Zeng+2013,Zeng+2019}, 
but this single quantity alone cannot uniquely constrain the composition \citep[e.g.][]{Seager+2007,Rogers&Seager10a, Dorn+15}.
For close-in super-Earths, recent studies predict their compositions based on the observed planetary radius distribution \citep[e.g.,][]{Lopez17,Owen&Wu17}, but the compositions inferred this way depend on the assumption about how the radius distribution is sculptured  \citep{Owen&Adams19}.

So-called disintegrating planets discovered by the {\it Kepler}/{\it K2} missions \citep{Rappaport+2012, Rappaport+2014, Sanchis-Ojeda+2015} could offer a unique opportunity to investigate exoplanets' solid compositions. Disintegrating planets are ultra-short-period exoplanets that exhibit asymmetric transit light curves. 
An example is K2-22b, which is a planet orbiting close to an M-type star and exhibits asymmetric and highly time-variable transit curves \citep{Sanchis-Ojeda+2015}.
Disintegrating planets are thought to have an evaporating solid surface that produces mineral vapor, and their asymmetric light curves are interpreted as being due to a comet-like tail forming from the recondensing vapor \citep[e.g.,][]{Rappaport+2012}.
Models suggest that they are presumably solid planets of Moon to Mercury masses \citep[e.g.,][]{Rappaport+2012, Perez+2013}.
Therefore, by performing transmission spectroscopy of the dust tails, we may be able to directly probe the solid composition of low-mass planets.
Although there are only a handful of known candidates for disintegrating planets \citep{Rappaport+2012,Rappaport+2014,Sanchis-Ojeda+2015,Jones+2020}, current and future missions for transit searches will discover more candidates.

\citet{Bodman+2018} explored the possibility of inferring the composition of the dust tails of disintegrating planets from  
 infrared observations with the {\it James Webb Space Telescope} (JWST). 
They found that silicate resonant features near 10 $\mu$m can
produce transit depths that are at least as large as those in the visible, and that
most of the features will be detectable with the JWST mid-infrared instrument (MIRI) for the particular case of K2-22b.

While JWST will serve as a powerful tool to observe silicate features in the wavelength range of 5--14 $\mu$m, silicates also produce a number of absorption features at longer wavelengths
(\citealt{Kitzmann+2018}; see also Figure~\ref{fig:g-global} of this paper). 
Therefore, transmission spectroscopy of the dust tails at longer wavelengths should be useful to put stronger constraints on what material the dust tails of disintegrating planets are made of. The proposed mission, {\it Space Infrared Telescope for Cosmology and Astrophysics} (SPICA), can potentially make such observations possible.
SPICA is a candidate space mission envisaged to be launched in the late 2020s or early 2030s \citep{2018PASA...35...30R}. 
Its mid-infrared instrument (SMI) enables low- to high-resolution spectroscopy in the wavelength range of 12--36 $\micron$ \citep{Kaneda+2016}, which is ideally complementary to spectroscopy with JWST.

In this paper, we investigate the feasibility of constraining the composition of disintegrating planets from combined transmission spectra obtained with future JWST and SPICA observations.
We construct a simple model for the spatial distribution
of the dust tails of disintegrating planets to produce their theoretical transmission spectra.
Since the solid compositions of exoplanets may be diverse \citep{Bond+2010, Jontof-Hutter+2019}, 
we consider various minerals that potentially constitute the dust tails.
Based on the expected capability of JWST and SPICA, we quantify the detectability of the absorption features of various minerals at infrared wavelengths as a function of the optical transit depth and distance to the system.
We show that combined transmission spectra from JWST and SPICA can place stronger constraints on the composition of disintegrating planets than spectra from JWST alone.

The organization of this paper is as follows.
In Section \ref{sec:model}, we describe our model for the spatial distribution and mineral composition of the dust tails.
In Section \ref{sec:th-spectra}, we present theoretical infrared transmission spectra of the tails in the limit of zero observational noise to identify the absorption features of the candidate minerals in the wavelength coverages of the JWST and SPICA mid-infrared instruments. 
In Section \ref{sec:detectability}, we account for expected  observational noise to quantify the detectability of mineral features as a function of the transit depth and distance to the planet. 
In Section \ref{sec:discussion}, we discuss some uncertainties of the tail model and possible applications of the proposed tail transmission spectroscopy to planet formation studies.
We summarize our findings in Section \ref{sec:conclusion}.

\section{Model} \label{sec:model}

\subsection{Transit Model 
} \label{subsec:transit}

\begin{figure}[t]
\includegraphics[bb=0.000000 0.000000 258.000000 128.000000, width=8.5cm]{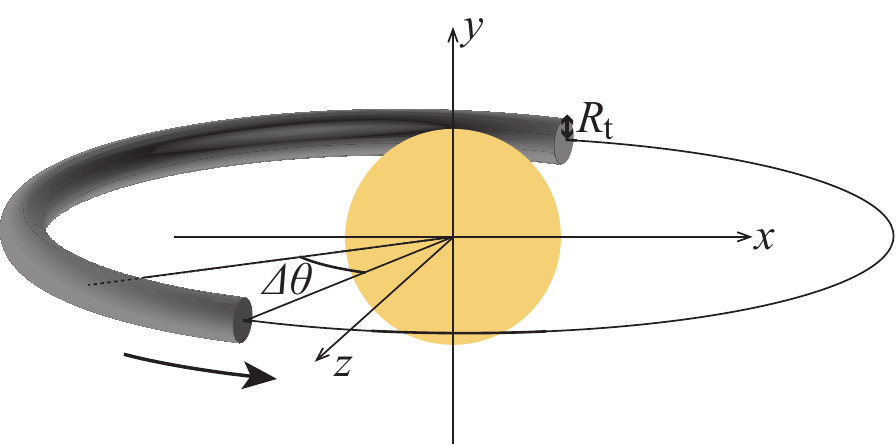}
\caption{
Schematic illustrating showing the geometry of the dust tail. The tail is approximated by a curved cylinder of thickness $R_{\rm t}$ orbiting the central star. The z-axis is in the direction of the observer, and the y-axis is perpendicular to the orbital plane of the dust tail. 
}
\label{fig:tail-model}
\end{figure}

We use a simple analytic model for the spatial distribution of the dust tails.
Figure \ref{fig:tail-model} schematically shows the geometry of a modeled tail.
The dust tail is approximated by a cylinder of radius $R_{\rm t}$, trailing with the planet
along a circular orbit at distance $a$ from the central star.
 We employ such a simple model because the dust tail is expected to be largely optically thin (see below), for which a time-integrated transmission depth is nearly proportional the total amount of dust grains passing in front of the central star.
For the same reason, we neglect a leading tail that could be associated with a real disintegrating planet, including K2-22b \citep{Sanchis-Ojeda+2015}. As we show at the end of this section, our simplified model can reasonably reproduce the transit light curve of K2-22b except for its ingress.

For simplicity, we assume that the line of sight of an observer lies on the orbital plane of the planet and dust tail.
We introduce a Cartesian coordinate system centered on the star with the $z$ and $y$ axes taken to be in the direction of the observer and perpendicular to the orbital plane, respectively (see Figure \ref{fig:tail-model}).
The transit depth $\delta$ at a given time and  a wavelength $\lambda$ is then given by
\begin{equation}
\delta(\lambda)= \frac{\int _{\sqrt{x^2+y^2} \leq R_{\rm star}}
(1-e^{-\tau(x,y; \lambda)}) dxdy
}{\pi R_{\rm star}^{2}},    \label{eq: transit-d}
\end{equation}
where $R_{\rm star}$ is the radius of the central star, $\tau$ is the optical depth along the line of sight of the observer. 
We applied our dust tail model to the observed light curve for the disintegrating planet K2-22b \citep{Sanchis-Ojeda+2015} and found that the model reproduces the
half-width of the light curve at an accuracy of $\la 40$ \% when the maximum transit depth of our model is set equal to the observed depth.

The optical depth $\tau$ is given by the product of the number density and extinction cross section of the dust grains integrated over the observer's line of sight. Following \citet{Rappaport+2014}, we simply assume that the  tail decays exponentially with the angular distance from the planet, $\Delta \theta$ (see Figure \ref{fig:tail-model} for the definition of $\Delta \theta$), expressing $\tau$ as
\begin{equation}
\tau(x,y; \lambda) =  \int  
\sigma_{\rm ext}(\lambda) n_0
e^{-\alpha \Delta\theta(x,z)} dz,
\label{eq:tau}
\end{equation}
where $\sigma_{\rm ext}$ and $n_{0}$ are the extinction cross section and number density of the grains at $\Delta \theta  =0$, respectively, and $\alpha (>0)$ is the factor that characterize the length scale over which the tail decays.
The exponential function in Equation~\eqref{eq:tau} encapsulate the net effect of the tail's decay due to grain sublimation on the grain number density and on the grain size. This simple prescription, although somewhat ad hoc, allows us to take $\sigma_{\rm ext}(\lambda)$ as constant in space (see Appendix C of \citealt{Rappaport+2014} for a validation of this prescription).

The dust grains are approximated as compact spheres of equal radius $r_{\rm d}$, and their extinction cross section is calculated  using the publicly available Mie calculation code LX-MIE \citep[][]{Kitzmann+2018}.
The complex refractive indices of the minerals constituting the grains are taken from literature date (see Section \ref{subsec:comp} for details).
The typical grain size in the dust tails is observationally inferred to be in the range 0.1--1 $\micron$ from the forward scattering of the tails and from the wavelength dependence of the transit depths \citep{Budaj2013, Rappaport+2014, Sanchis-Ojeda+2015}.
For simplicity, we fix $r_{\rm d} = 1~\micron$ in our calculations.

\begin{figure}
\hspace{-4mm}
\includegraphics[bb=0.000000 0.000000 360.000000 252.000000, scale=0.8]{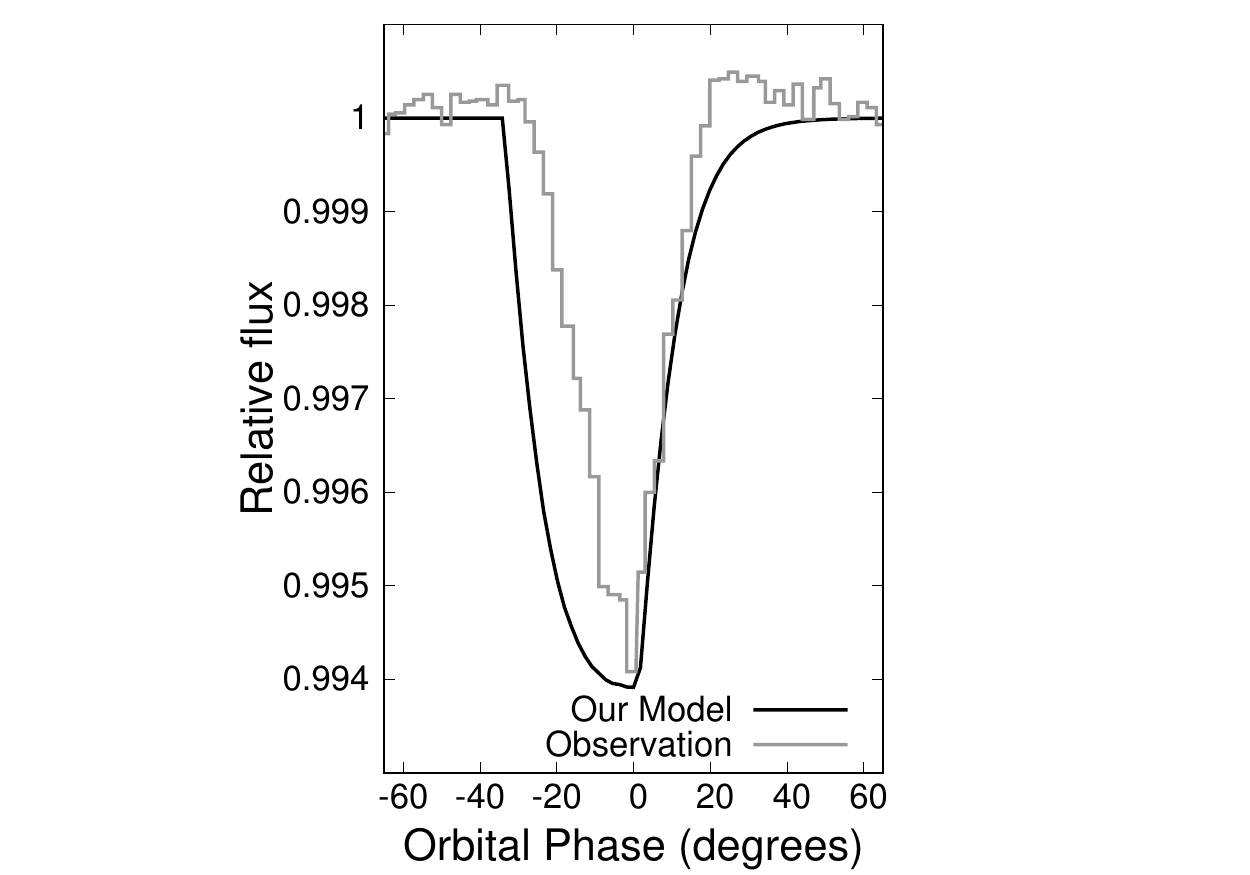}
\caption{Transit light curve for a dust tail composed of SiO$_{2}$ grains from the tail model with $R_{\rm t}=0.02R_{\rm star}$, $n_{0}= 5.6 \times 10^{4}~\rm m^{-3}$, and $\alpha = 1/ (0.133$ rad). For comparison, the gray line shows an averaged observed light curve for K2-22b, taken from Figure 6 of \citet{Sanchis-Ojeda+2015} using WebPlotDigitzer (\url{https://automeris.io/WebPlotDigitizer/}).
}
\label{fig:lc-fid}
\end{figure}

Our model involves three important parameters:  $R_{\rm t}$, $n_{0}$, and $\alpha$.
We choose the parameter values so that the dust tail is optically thin at the wavelength of the {\it Kepler} band ($\lambda_{\rm Kepler} =$ 0.42--0.9 $\mu$m).
If the dust surrounding an evaporating planet were optically thick, starlight would not reach the planet's surface and evaporation would be quenched \citep{Rappaport+2012, Perez+2013, Lieshout+2016}. 
Figure \ref{fig:lc-fid} illustrates how a transit curve from our model compares with a real transmission light curve for the disintegrating planet K2-22b \citep{Sanchis-Ojeda+2015}. 
For $R_{\rm t}=0.02R_{\rm star}$ and $n_{\rm 0}=5.6\times{10}^4~{\rm m}^{-3}$, the dust tail is sufficiently optically thin and also reproduces the maximum transit depth. The slope of the light curve in the transit egress (corresponding to orbital phase $\approx 0$--20$^\circ$) is best fit with the choice $\alpha=1/(0.133~{\rm rad})$. 
 Since our model neglects a leading tail, the ingress shows a less good match than the engress.
In Section~\ref{sec:th-spectra}, we adopt this parameter set to produce noise-free theoretical transmission spectra.

Because the grain extinction cross section depends on wavelength, transit light curves at different wavelengths exhibit different transit depths.
In this study, we define transmission spectra as the maximum transit depth as a function of wavelength.

\subsection{Dust Composition} \label{subsec:comp}

\begin{figure*}[t]
    \centering
    \includegraphics[bb=0.000000 0.000000 1292.000000 521.000000, width=0.99\textwidth]{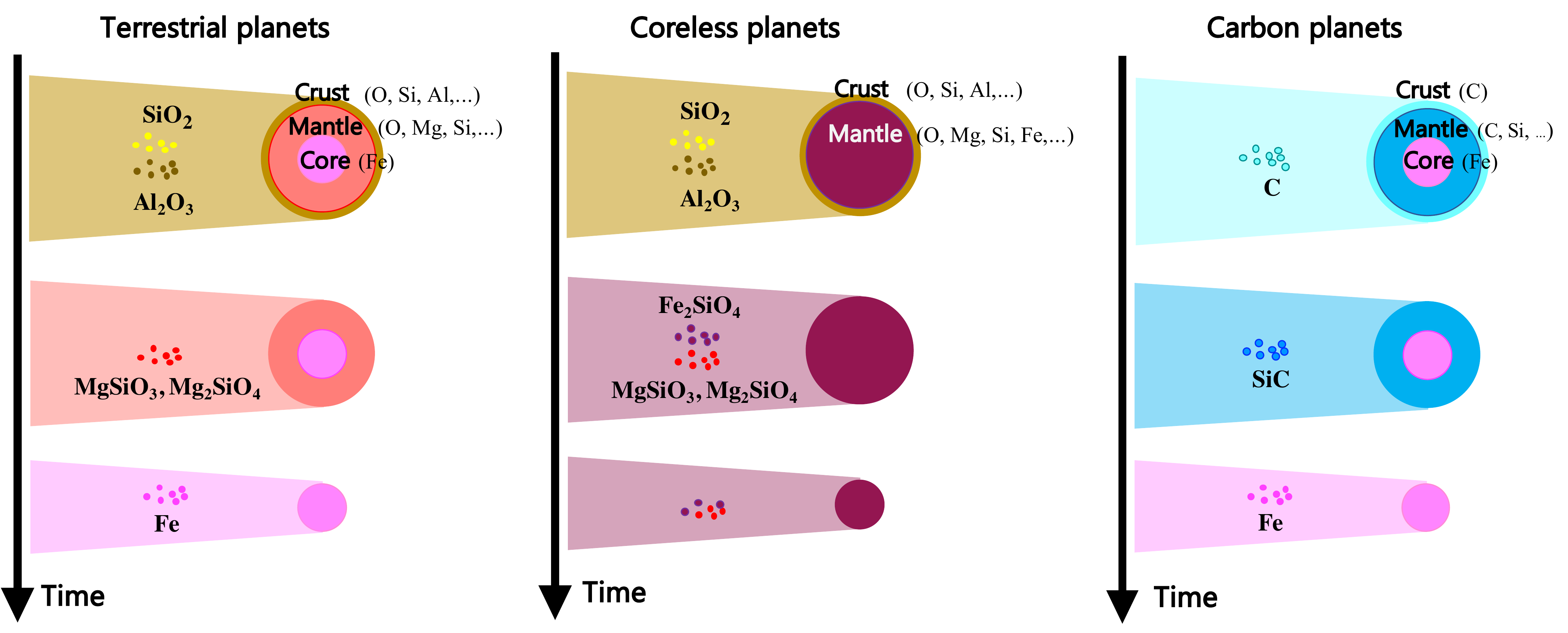}
    \caption{
    Schematic illustration of candidate compositions of the dust tail generated by the evaporation of terrestrial, coreless \citep{Elkins-Tanton+2008}, and carbon \citep{Kuchner&Seager2005} exoplanets (see also Section~\ref{subsec:links}).
    The time evolution of dust compositions goes along the arrows if the planets evaporate (disintegrate) from their crusts to their cores.}
    \label{fig:tail-comp}
\end{figure*}

We consider 8 candidate minerals for the dust tails of terrestrial and non-terrestrial planets (Figure~\ref{fig:tail-comp} and Table \ref{tab:minerals}). 
By terrestrial planets we mean rocky planets comprised of a crust containing O, Si, and Al; a mantle containing O, Mg, and Si; and an iron core. The vapor from the crust, mantle, and core is assumed to form dust tails made of SiO$_2$ (silica) and Al$_{2}$O$_{3}$ (alminium oxide),  MgSiO$_3$ (enstatite) and Mg$_2$SiO$_4$ (forsterite), and pure Fe, respectively.
Coreless planets are hypothetical planets that have no metallic core but have a giant silicate mantle that is abundant in iron \citep{Elkins-Tanton+2008}. The vapor from the mantle of coreless planets would condense into Mg- and Fe-bearing silicates, which we model with MgSiO$_3$, Mg$_2$SiO$_4$, and Fe$_2$SiO$_4$ (fayalite).
Carbon planets are solid planets with a carbon-to-oxygen ratio (C/O) larger than 1. 
The vapor from this type of planets would condense into 
SiC (silicon carbide) and graphite 
\citep[e.g.,][]{ Woitke+2018}.
In Section~\ref{subsec:links} we discuss in more detail how the mineral composition of dust tails can be linked to the composition of the disintegrating planets themselves.
For simplicity, our synthetic transmission spectra assume that the dust tail is composed of a single mineral species.

Computation of the grain extinction cross section requires data of the complex refractive index as a function of wavelength.
In principle, the refractive index differs between crystalline or amorphous materials, and Table \ref{tab:minerals} summarizes which material is considered for each mineral (``cryst.''~and ``amo.''~refer to crystalline and amorphous materials, respectively).
The optical constants for amorphous silicates are similar (see Figure 1 of \citealt{Kitzmann+2018}), so we arbitrarily take amorphous Mg$_2$SiO$_4$ as a representative amorphous silicate.
The refractive indices for crystalline Mg$_2$SiO$_4$ and MgSiO$_3$ are taken from  \citet{Fabian+2001} and  \citet{Jaeger+1998}, respectively.
The refractive indices for the other minerals 
are taken from the compilation by \citet{Kitzmann+2018}.

\begin{deluxetable}{ccc}[t]
\tablecaption{Candidate Minerals for the Dust Tails of Disintegrating Planets \label{tab:minerals}}
\tablecolumns{5}
\tablewidth{0pt}
\tablehead{
\colhead{Mineral} &
\colhead{Planet type} &
\colhead{Source part} 
}
\startdata
SiO$_2$ (amo.)  & Terrestrial & Crust \\
Al$_2$O$_3$ (amo.) & Terrestrial & Crust \\
MgSiO$_3$ (cryst.) & Terrestrial/Coreless  & Mantle \\
Mg$_2$SiO$_4$ (cryst. \& amo.) & Terrestrial/Coreless & Mantle \\
Fe$_2$SiO$_4$ (cryst.) & Coreless  & Mantle \\
Fe (metallic iron)  & Terrestrial/Carbon & Core  \\
C (graphite) & Carbon & Crust  \\
SiC (cryst.) & Carbon & Mantle  \\
\enddata
\end{deluxetable}

\section{Theoretical Transmission Spectra \label{sec:th-spectra}} 
In this section, we present theoretical transmission spectra of the dust tails in the ideal limit of no observational noise and identify spectral features of candidate minerals in the JWST and SPICA wavelength ranges.

\begin{figure*}[t]
 \vspace{2mm}
 \begin{minipage}[b]{0.33\linewidth}
  \centering
  \includegraphics[bb=0.000000 0.000000 360.000000 252.000000, scale=0.48]
  {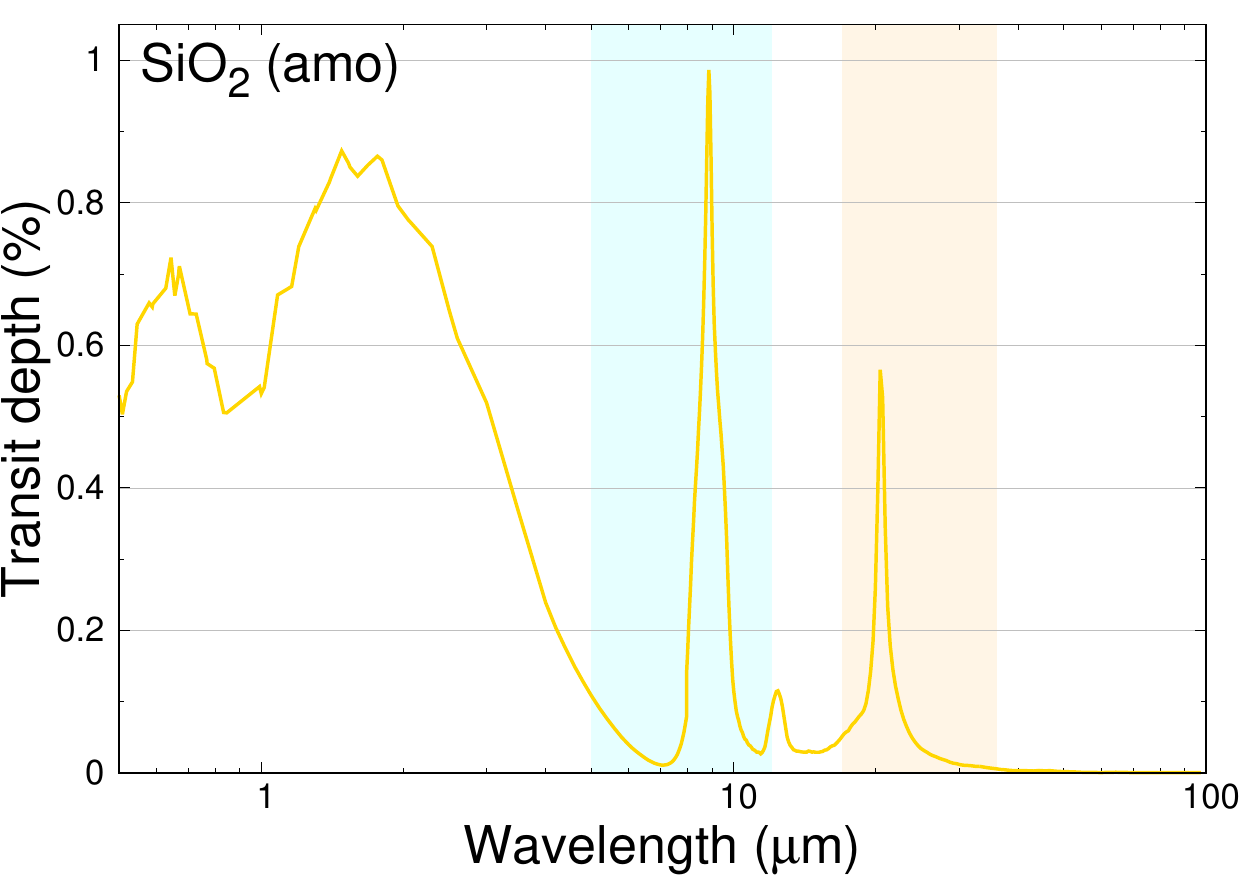}
 \end{minipage} 
 \begin{minipage}[b]{0.33\linewidth}
  \centering
  \includegraphics[bb=0.000000 0.000000 360.000000 252.000000, scale=0.48]
  {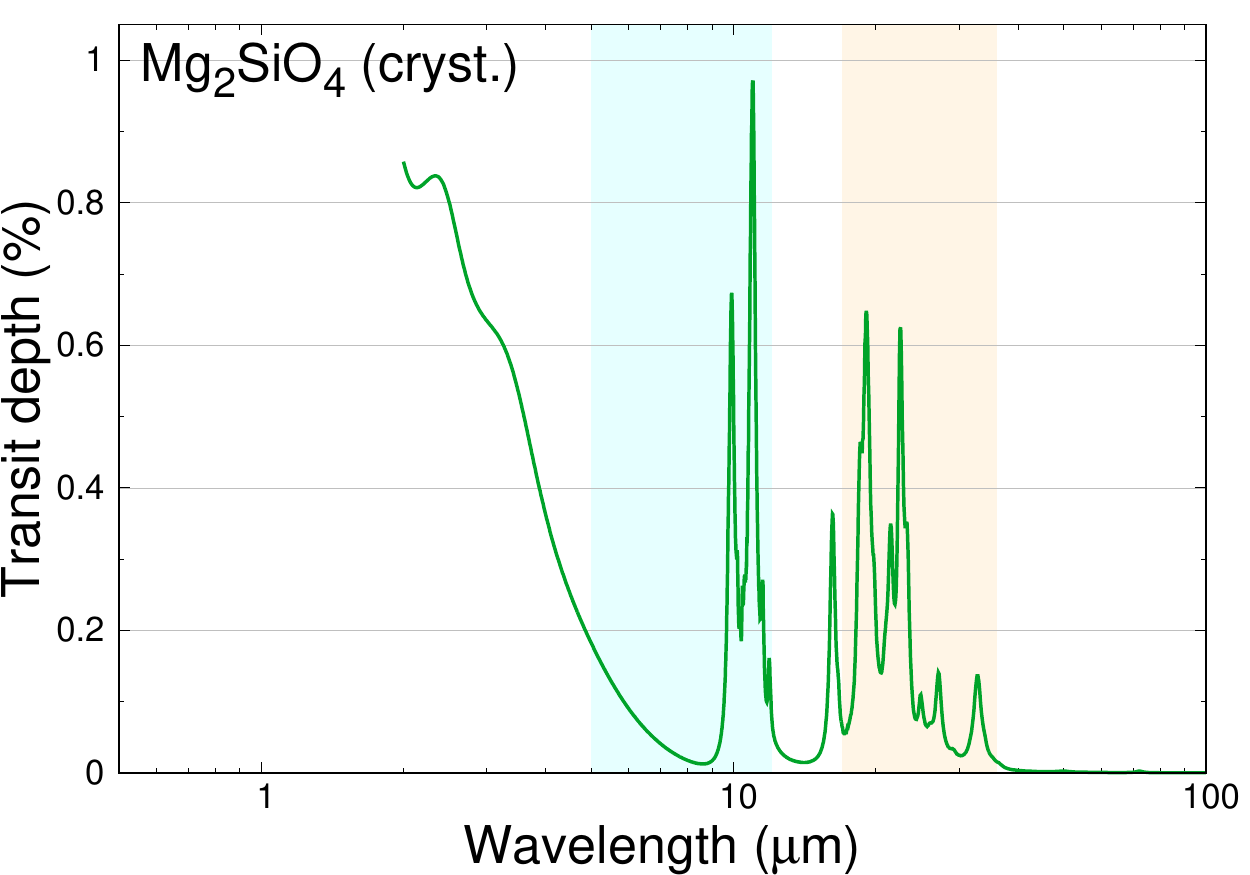}
 \end{minipage} 
 \begin{minipage}[b]{0.33\linewidth}
  \centering
  \includegraphics[bb=0.000000 0.000000 360.000000 252.000000, scale=0.48]
  {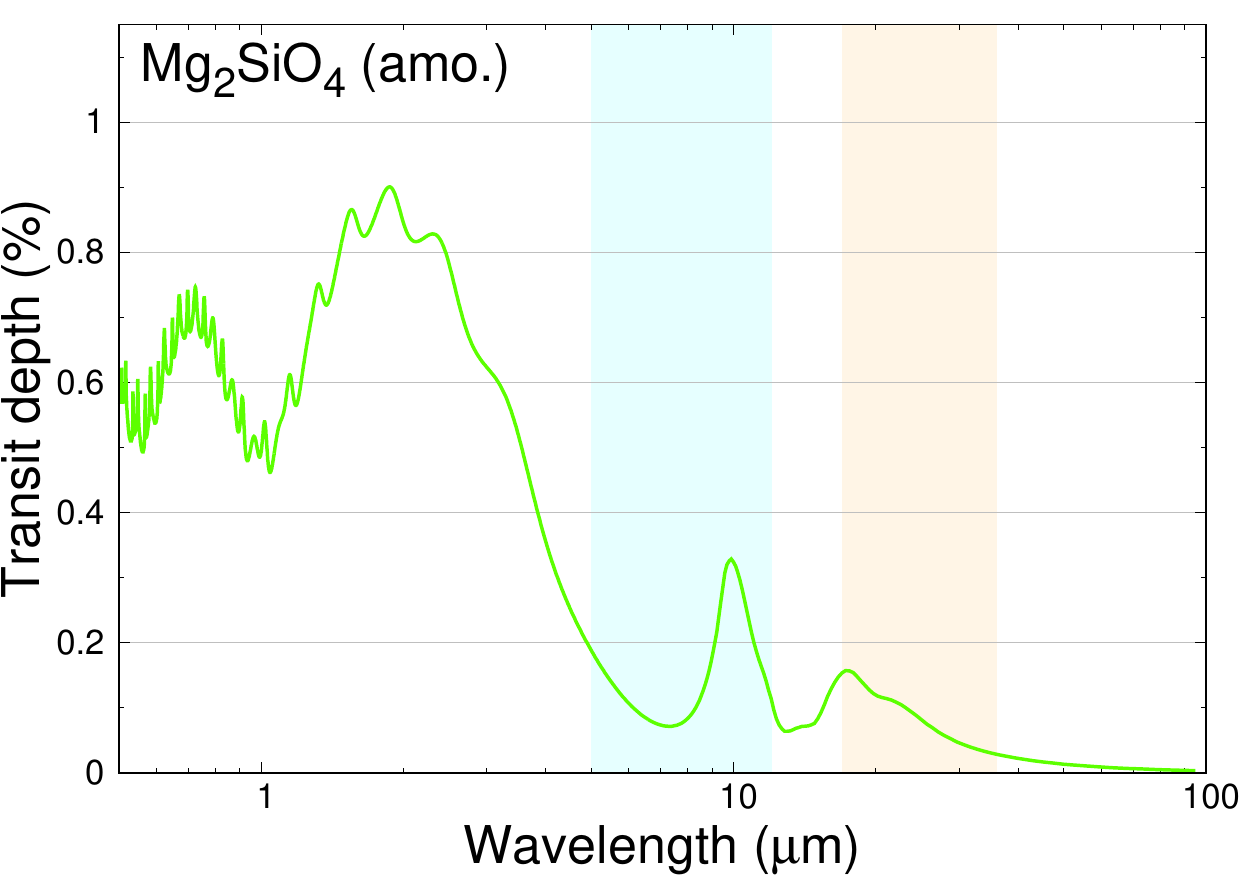}
 \end{minipage} \\
  \begin{minipage}[b]{0.33\linewidth}
  \centering
  \includegraphics[bb=0.000000 0.000000 360.000000 252.000000, scale=0.48]
  {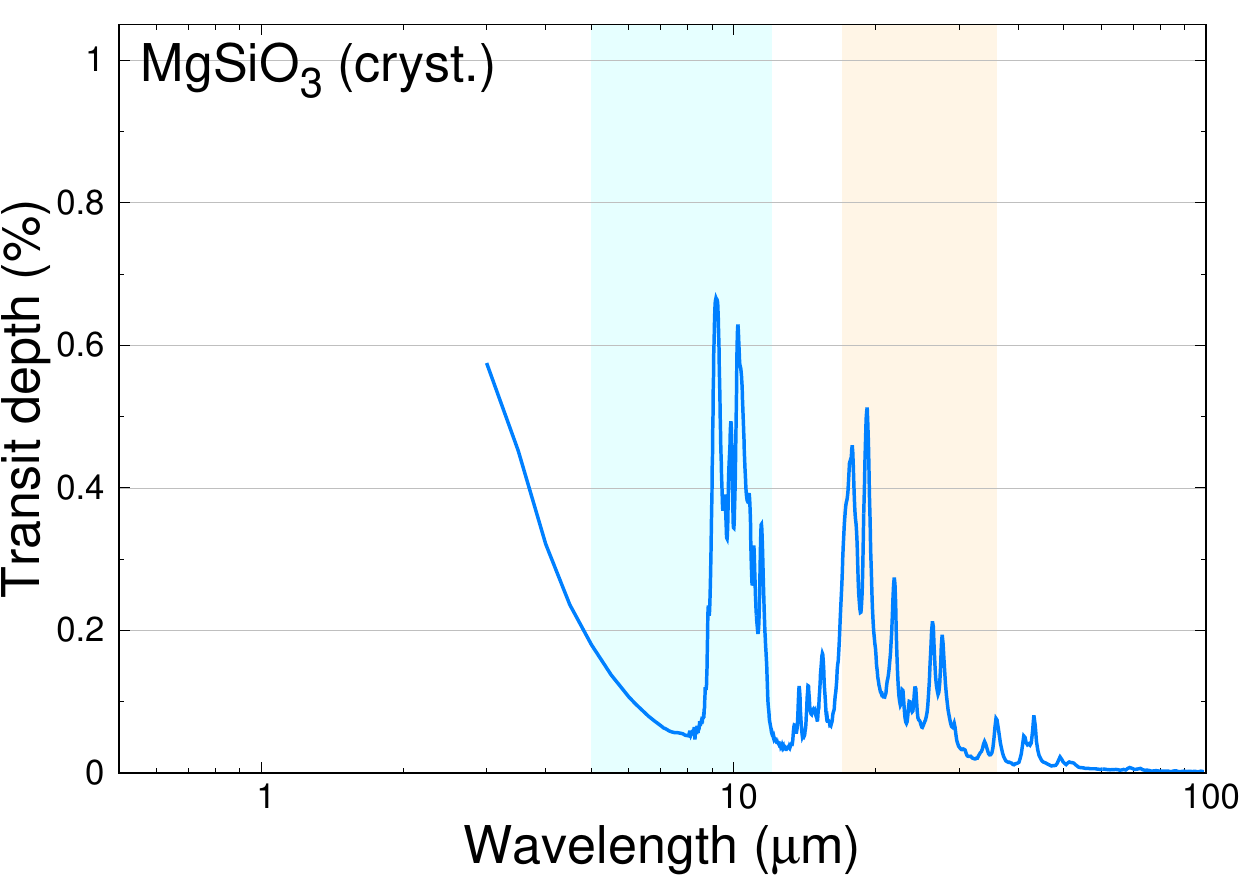}
 \end{minipage} 
 \begin{minipage}[b]{0.33\linewidth}
  \centering
  \includegraphics[bb=0.000000 0.000000 360.000000 252.000000, scale=0.48]
  {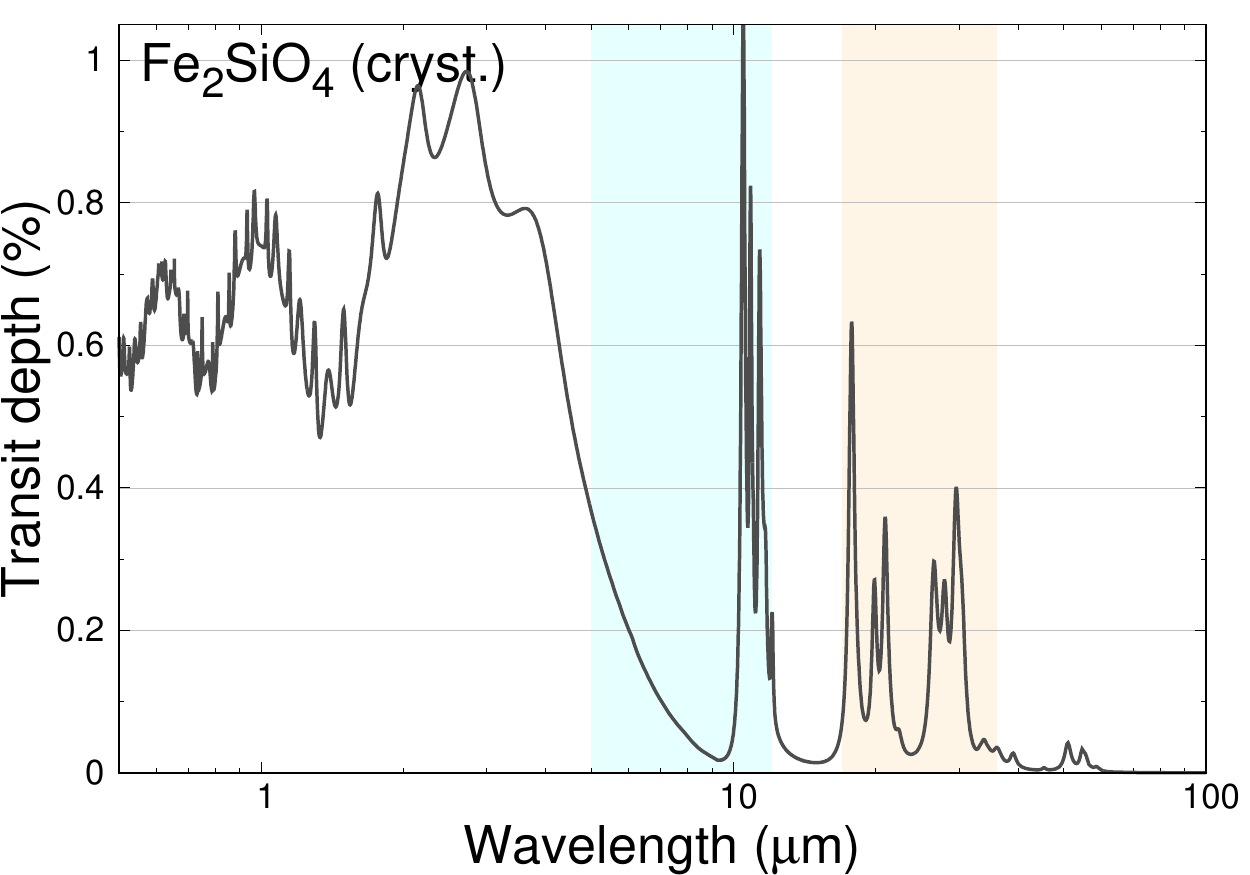}
 \end{minipage} 
  \begin{minipage}[b]{0.33\linewidth}
  \centering
  \includegraphics[bb=0.000000 0.000000 360.000000 252.000000, scale=0.48]
  {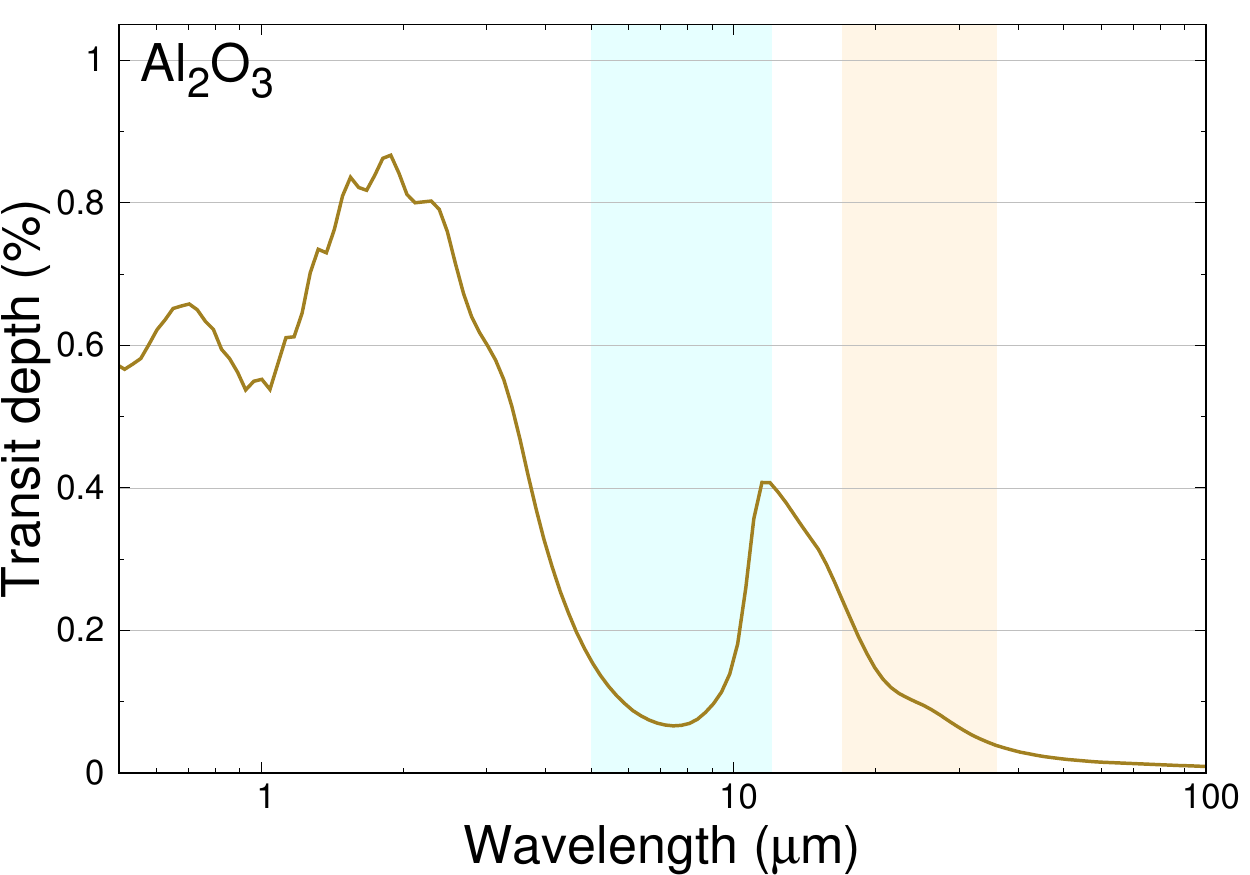} 
 \end{minipage} \\
 \begin{minipage}[b]{0.33\linewidth}
  \centering
  \includegraphics[bb=0.000000 0.000000 360.000000 252.000000, scale=0.48]
  {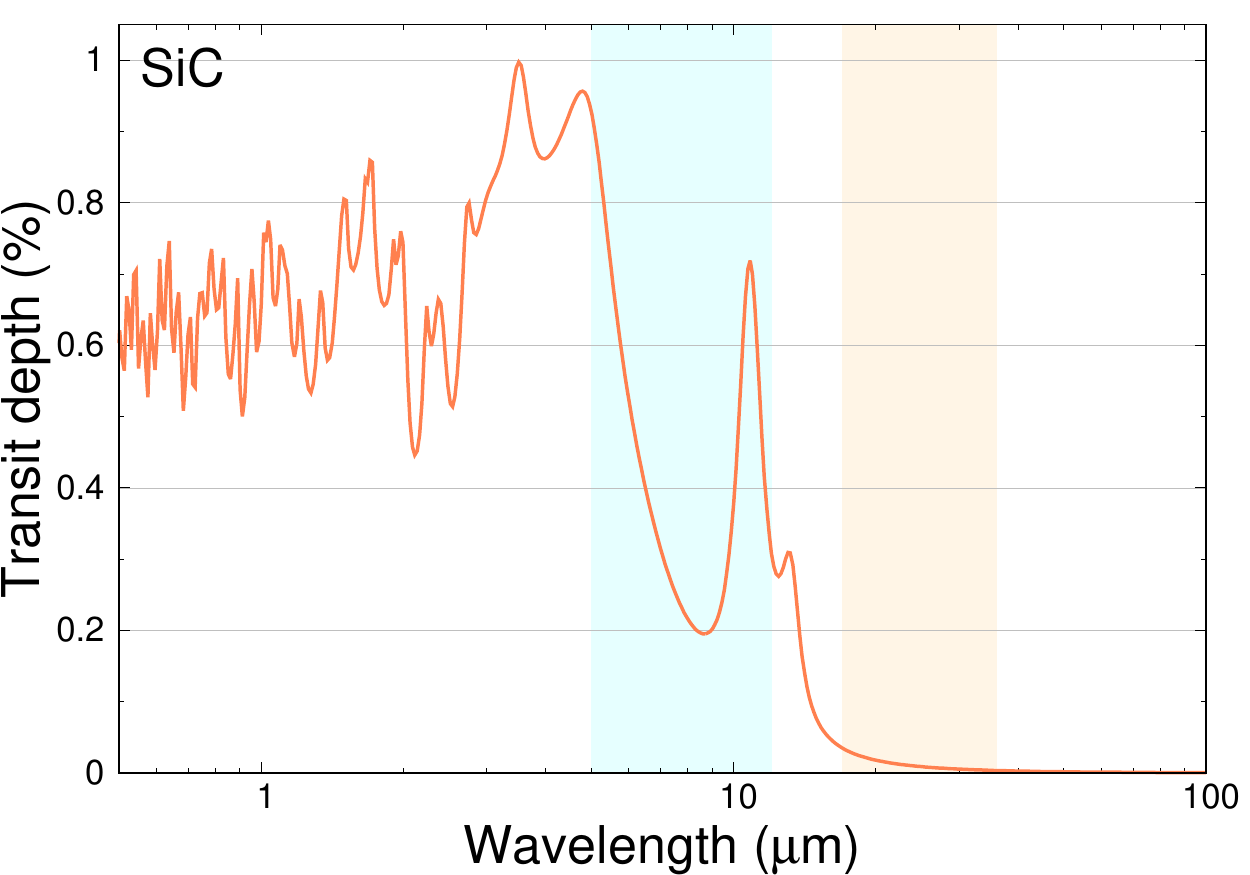}
 \end{minipage} 
 \begin{minipage}[b]{0.33\linewidth}
  \centering
  \includegraphics[bb=0.000000 0.000000 360.000000 252.000000, scale=0.48]
  {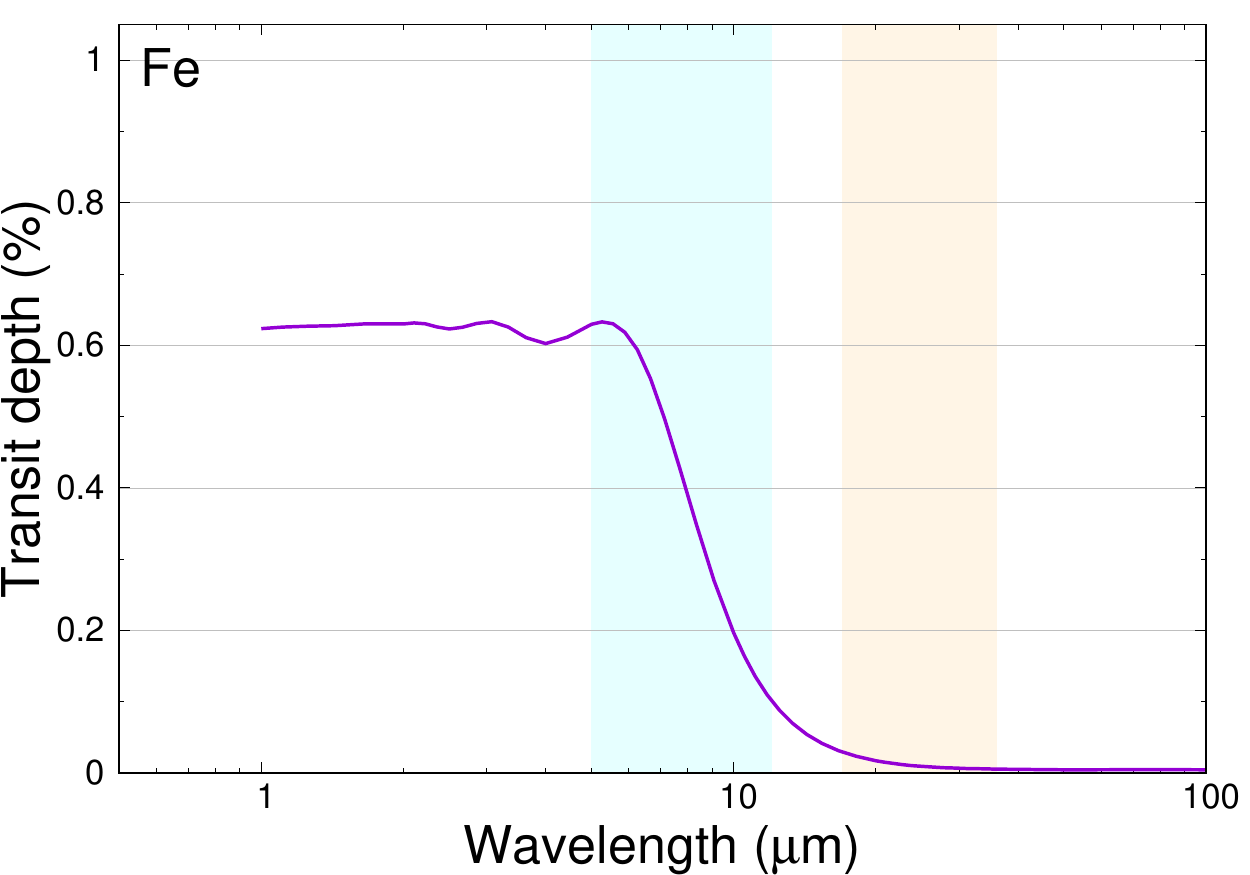}
 \end{minipage} 
 \begin{minipage}[b]{0.33\linewidth}
  \centering
  \includegraphics[bb=0.000000 0.000000 360.000000 252.000000, scale=0.48]
  {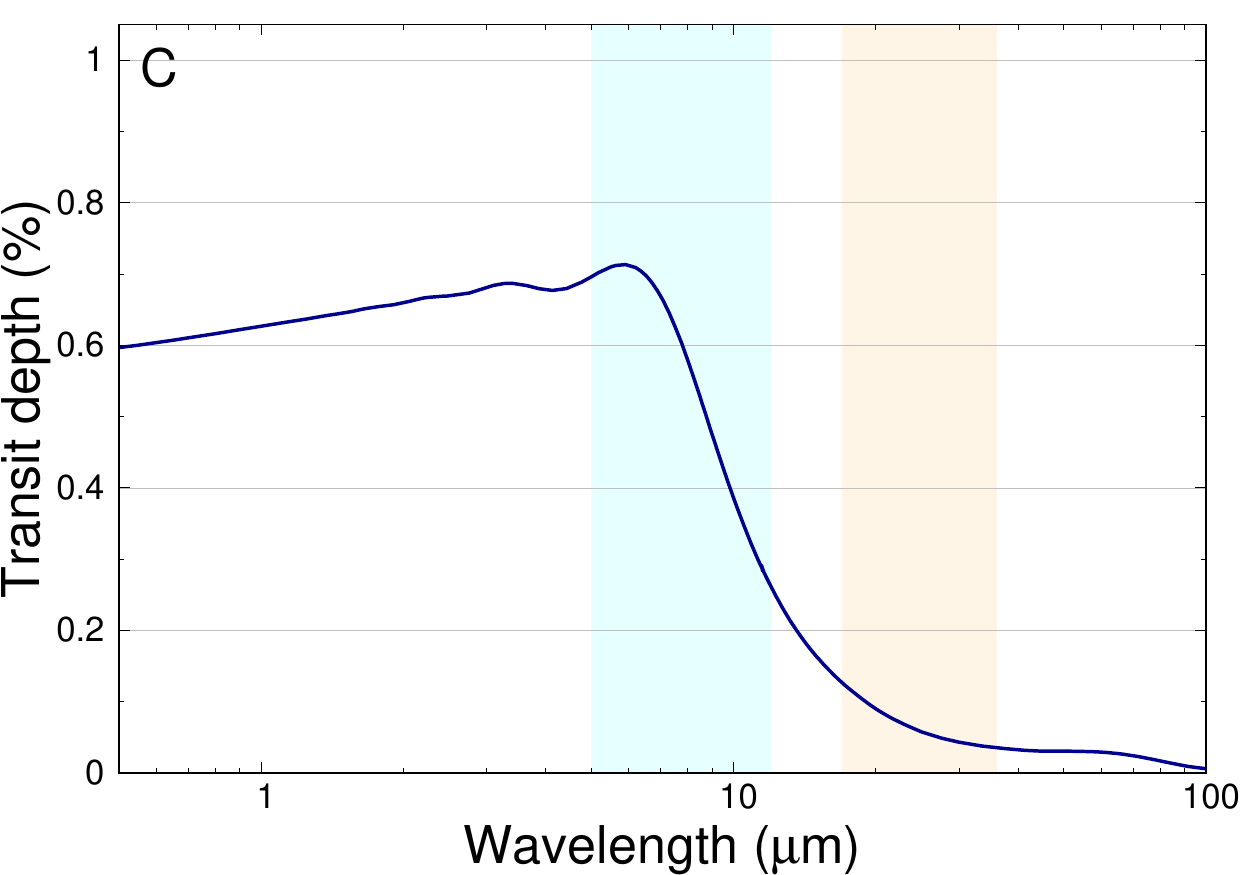}
 \end{minipage} \\
 \vspace{-2mm}
\caption{Theoretical transmission spectra for different minerals from the fiducial tail model.  Here, ``cryst.'' and ``amo.'' refer to crystalline and amorphous materials, respectively.
The cyan and orange shaded regions mark wavelength ranges covered by JWST MIRI/LRS (5--12 $\mu$m) and SPICA SMI/LR (17--36 $\mu$m), respectively.
}
\label{fig:g-global}
\end{figure*}

Figure~\ref{fig:g-global} shows the theoretical infrared transmission spectra from the fiducial tail model for all candidate minerals considered in this study. A variety of absorption features are visible at wavelengths longer than 5 $\micron$, and some of them fit in the wavelength ranges covered by JWST MIRI and SPICA SMI.
The wiggles appearing in the spectrum curves at $\lambda \la 5~\rm \micron$ are so-called interference and ripple structures characteristic of Mie scattering by spheres with $r_{\rm d} \ga \lambda/2\pi$ \citep[see, e.g.,][]{Bohren&Huffman83}. These structures largely depend on $r_{\rm d}$ and hence are not purely inherited from the bulk optical properties of the materials. 
For this reason, below we focus on the absorption features at $\lambda > 5~\rm \micron$.

\begin{deluxetable*}{cccc}[t]
\tablecaption{Classification of the Candidate Minerals Based on the Position of Absorption Peaks}
  \label{table:group}
\tablecolumns{4}
\tablewidth{0pt}
\tablehead{
\colhead{Group} &
\colhead{Peaks/Features with JWST} &
\colhead{Peaks/Features with SPICA} & 
\colhead{Minerals}
}
\startdata
    (a) & \checkmark  & \checkmark  & SiO$_{2}$, Fe$_{2}$SiO$_{4}$, Mg$_{2}$SiO$_{4}$, MgSiO$_{3}$ \\
     (b) & \checkmark  &   & Al$_{2}$O$_{3}$, SiC \\
      (c) &    &   & Fe, C \\
\enddata
\end{deluxetable*}

\begin{figure*}
 \begin{minipage}[b]{0.5\linewidth}
  \includegraphics[bb=0.000000 0.000000 360.000000 252.000000, scale=0.65]
  {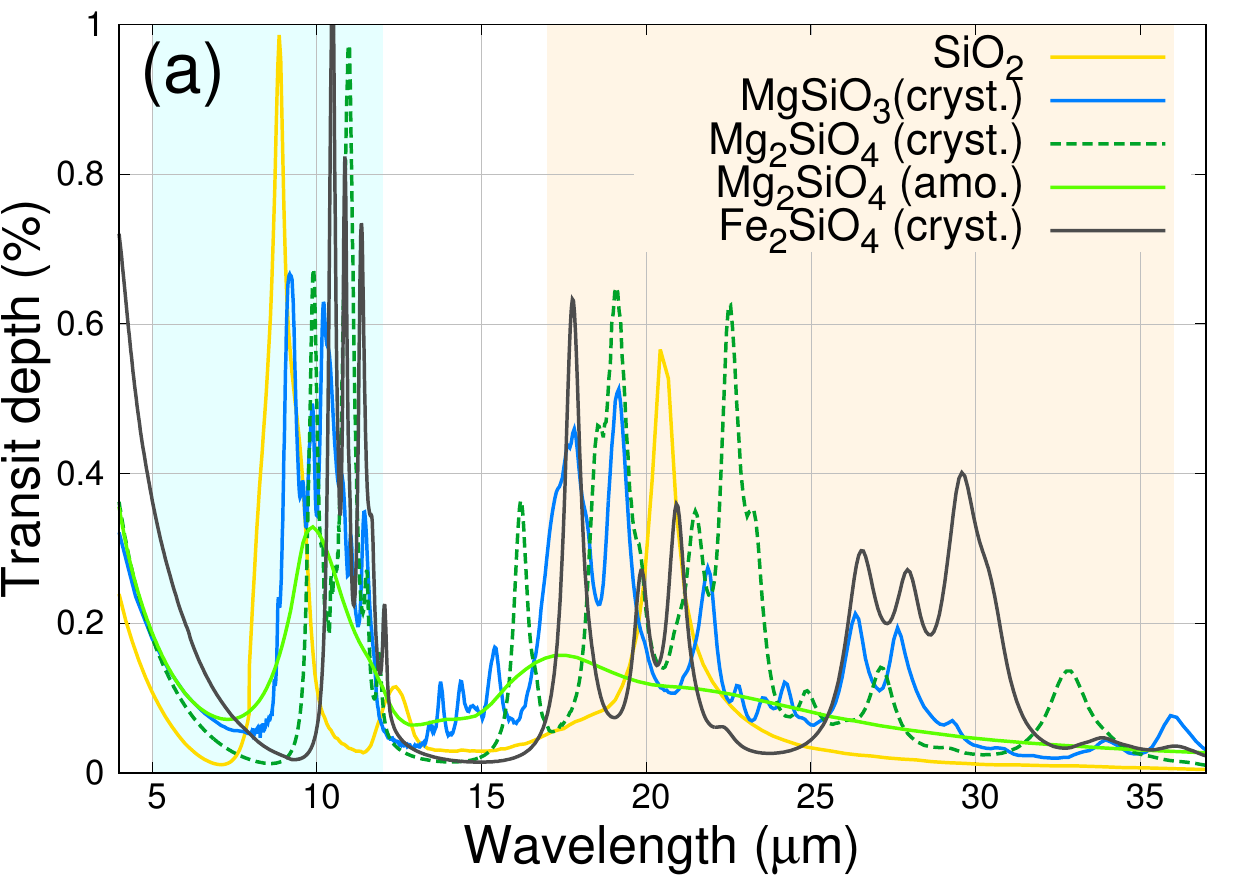}
 \end{minipage} 
 \begin{minipage}[b]{0.5\linewidth}
  \includegraphics[bb=0.000000 0.000000 360.000000 252.000000, scale=0.65]
  {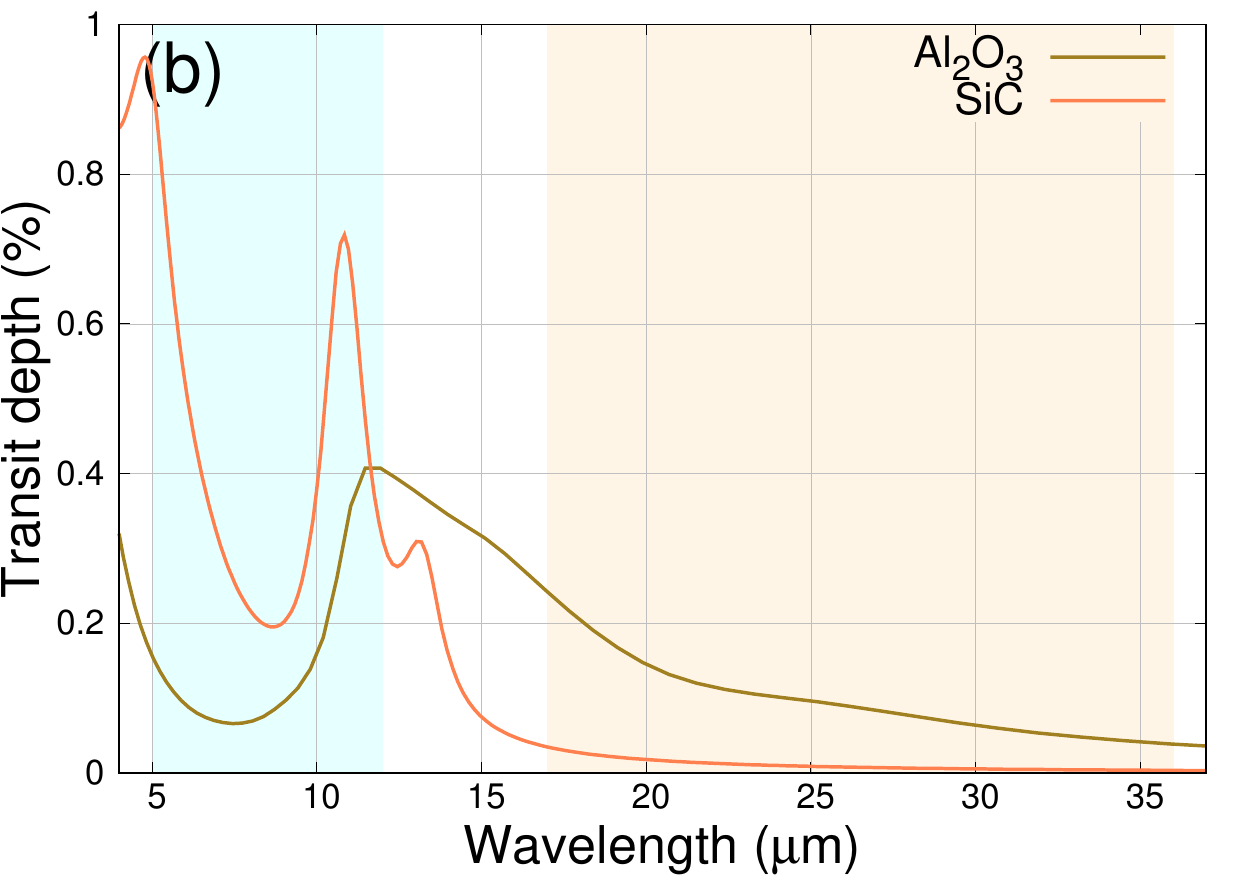}
 \end{minipage} \\
   \hspace{4mm}
 \begin{minipage}[b]{0.5\linewidth}
  \includegraphics[bb=0.000000 0.000000 360.000000 252.000000, scale=0.65]
  {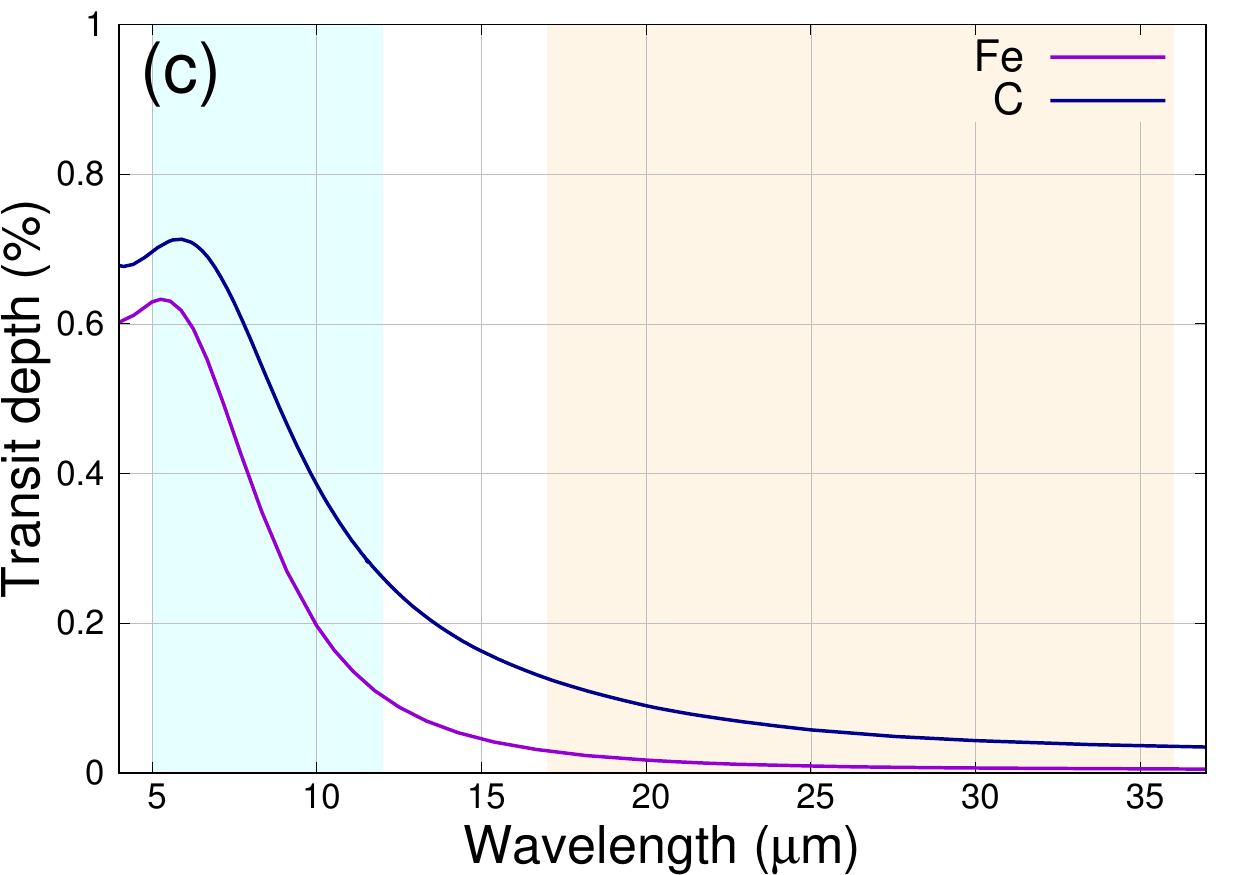}
 \end{minipage} 

\caption{Theoretical transmission spectra from the fiducial tail model classified into three groups (a)--(c) based on the presence or absence of features in the JWST MIRI/LRS and SPICA SMI/LR wavelength ranges (see Table \ref{table:group} for the classification). The absorption features seen at $\lambda \approx 5~\micron$ are the edges of the interference structure of the $1~\micron$-sized spherical particles and do not directly reflect the intrinsic bulk absorption properties of the minerals (see text). 
\label{fig:th-spectra}}
\end{figure*}

Based on the presence or absence of absorption peaks/features in the wavelength ranges covered by JWST and SPICA, the candidate minerals can be crudely classified into three groups as summarized in Table \ref{table:group}  (see also Figure \ref{fig:th-spectra}).
Group (a) includes SiO$_2$ and silicates (Fe$_{2}$SiO$_{4}$, Mg$_{2}$SiO$_{4}$, and MgSiO$_{3}$), which show absorption peaks both in the JWST and SPICA wavelength regions. 
Group (b) consists of Al$_2$O$_3$ and SiC, which have an absorption peak at $\lambda \approx 10$--$12\micron$ in the JWST range but not in the SPICA range; 
Group (c) consists of Fe metal and graphite, which are featureless and only exhibit slopes in both wavelength ranges. 
The above classification already demonstrates the strength of combined transmission spectroscopy with JWST and SPICA in discriminating between the tail composition.

The spectral features of the individual minerals can be characterized as follows (see also Figure~\ref{fig:th-spectra}). 
For silicates, the peaks in the JWST wavelength range are narrowly concentrated at $\sim 10~\micron$. When viewed at a low spectral resolution, these materials only exhibit a single peak around $10~\micron$ as we demonstrate in Figure~\ref{fig:obs-spectra-Rp} in the following section. 
However, in the SPICA wavelength range, their spectral features are more diverse; crystalline Mg-bearing silicates have prominent peaks at 19--23 $\micron$, while crystalline Fe$_2$SiO$_4$ exhibits additional strong features at 26--30 $\micron$. 
This diversity is visible even in low-resolution spectra (see Figure~\ref{fig:obs-spectra-Rp}). This implies that SPICA could potentially serve as a useful tool to discriminate between silicates in the dust tail. 
Amorphous silicates (represented by amorphous Mg$_2$SiO$_4$) have a peak at 17--18 $\micron$, but this lies at at the edge of the SPICA wavelength coverage. This feature would appear as a slope rather than a peak in low-resolution spectra (see Section~\ref{sec:detectability}).
SiO$_2$ has a peak at 8.8 $\mu$m in the JWST range and at 20.5 $\mu$m in the SPICA range. The positions of the peaks are fairly insensitive to crystallinity because the refractive indices for silicate and amorphous SiO$_2$ are similar \citep{Kitzmann+2018}.
As for the minerals belonging to group (b), SiC has a peak at 10.8 $\mu$m, whereas Al$_{2}$O$_{3}$ has a peak at 12 $\mu$m.
When viewed in the SPICA wavelength range, SiC produces very low transit depths, while Al$_{2}$O$_{3}$ produces a spectral slope whose amplitude is comparable to that of the feature in the JWST range. 
These differences might be useful for discriminating between the two minerals from transmission spectra. 
Obviously, the minerals belonging to group (c) would be  indistinguishable at mid-infrared wavelengths.

\section{Detectability of mineral features with future observations}
\label{sec:detectability}

To predict if the spectral features identified in the previous section will be detectable in future observations with JWST and SPICA, we need to estimate how large observational noise will be added to real transmission spectra.
In this section, we quantify the detectability of the mineral features of disintegrating planets by taking into account expected observational noise.

Because signals in transmission spectra of low-mass exoplanets are generally faint, spectral resolution must be compromised to gain sensitivity. Because the spectral features of solid materials are broad compared to molecular lines, low-resolution ($R = \lambda/\Delta \lambda \sim 10$, where $\Delta \lambda$ is the spectral resolution) spectroscopy would suffice to spectrally resolve the features. 
For this reason, we assume observations using the low-resolution mid-infrared spectrometers on JWST (MIRI/LRS; $\lambda = 5$--12 $\mu$m, $R = 40$--200) and on SPICA (SMI/LR; $\lambda = $17--36 $\mu$m, $R= 50$--150).
Following \citet{Bodman+2018}, we also assume binning of observed spectra into coarser spectral bins of $R = 10$ to further enhance signal-to-noise ratios.  
As an example, Figure~\ref{fig:obs-spectra-Rp} shows theoretical
transmission spectra for some selected candidate minerals 
from the tail model used in Section~\ref{sec:th-spectra} (see Figure~\ref{fig:g-global}), but now binned to $R = 10$.
With this reduced spectral resolution, fine features are largely smeared out, but broad features are still visible both in the JWST and SPICA bands.

We adopt a transit duration of $t=0.8$ hr, which is equal to the transit duration of K2-22b \citep{Sanchis-Ojeda+2015}.
The detectability estimates shown below are for conservative, single transit observations. 
A higher signal-to-noise ratio may be achieved by averaging transmission spectra from multiple transit events.
Because disintegrating planets generally have very short orbital periods, multiple transit observations of the planets are relatively easy to conduct. 
However, interpretation of such averaged spectra would require caution when the tails is highly time-variable.

\subsection{Observational Noise for JWST and SPICA} \label{subsec:noise}

To begin with, we calculate the 1$\sigma$ noise levels $\sigma_{\rm N}$ expected for a single transit observation with JWST and SPICA.

For both JWST MIRI and SPICA SMI, the sensitivity is largely if not entirely determined by the shot noise on zodiacal light emission  \citep{Glasse+2015,Sakon+2016}, the level of which depends on where the target is on the sky.
The mid-infrared zodiacal emission near the ecliptic plane is $\approx 80/15 \approx 5$ times higher that that near the ecliptic pole \citep{Sakon+2016}.
Unless otherwise noted, we consider targets near the ecliptic pole. For targets near the ecliptic plane, 
we augment the noise levels at all wavelengths by the factor of 5.

For JWST MIRI/LRS, \citet{Glasse+2015} provide analytic fits to the 
expected 10$\sigma$ sensitivity for a spectrally unresolved point source and to the spectral resolution $\Delta \lambda$ (their Equations 17 and 18, respectively) for a $t = 10^4~\rm s$ observation as a function of wavelength.
The fit for the sensitivity, however, slightly underestimates the up-to-date expected sensitivity available online\footnote{\href{https://jwst-docs.stsci.edu/mid-infrared-instrument/miri-predicted-performance/miri-sensitivity}{MIRI Sensitivity - JWST User Documentation}}.  
We correct for the difference by multiplying the fit by a factor of $\exp[0.04(14-\lambda[\micron])^{1.5}]$.
We then compute the 1$\sigma$ noise level for the flux density by dividing the corrected 10$\sigma$ sensitivity for spectrally unresolved targets by $10\Delta \lambda$.   
We finally obtain the 1$\sigma$ noise level for $R = 10$ and $t = 0.8~\rm hr$ by assuming that the noise simply obeys Poisson statistics, i.e., $\sigma_{\rm N} \propto R^{1/2}t^{-1/2}$. 

For SPICA SMI, 
the expected 5$\sigma$ noise level for an 1-hr observation using SMI/LR, when scaled to $R = 80$, is $\approx 25~{\rm \mu Jy}$ according to the latest version of the SPICA SMI Fact Sheet\footnote{\url{https://www.ir.isas.jaxa.jp/SPICA/SPICA_HP/SMI/instrument-en.html}}.  
Again assuming Poisson noise, we obtain a 1$\sigma$ noise level of $\sigma_{\rm N} \approx 5(t/1~{\rm hr})^{-1/2}(R/80)^{1/2}~{\rm \mu Jy} \approx 2(t/0.8~{\rm hr})^{-1/2}(R/10)^{1/2}~{\rm \mu Jy}$.

To translate the noise in the flux density into the noise in the transit depth, one needs to assume the flux density of the emission from the central star.
Assuming blackbody radiation, the flux density
$F_{\rm star}$ of the starlight is given by 
\begin{equation}
F_{\rm star} (\lambda) = \frac{R_{\rm star}^{2}}{d^{2}} \pi B (\lambda, T_{\rm star}), \label{eq:E-star}  
\end{equation}
where $d$ is the distance from the Earth to the star, $T_{\rm eff}$ is the stellar effective temperature, and $B$ is the Planck function.
In this section, we fix $R_{\rm star}=0.55\,R_{\odot}$ and 
$T_{\rm star}=3800$ K, which apply to K2-22 \citep{Sanchis-Ojeda+2015}, but vary the value of $d$ to be able to predict the detectability of mineral features for as yet undiscovered disintegrating planets.

\begin{figure}
\vspace {2mm} 
\includegraphics[bb=0.000000 0.000000 360.000000 252.000000, width=\linewidth]{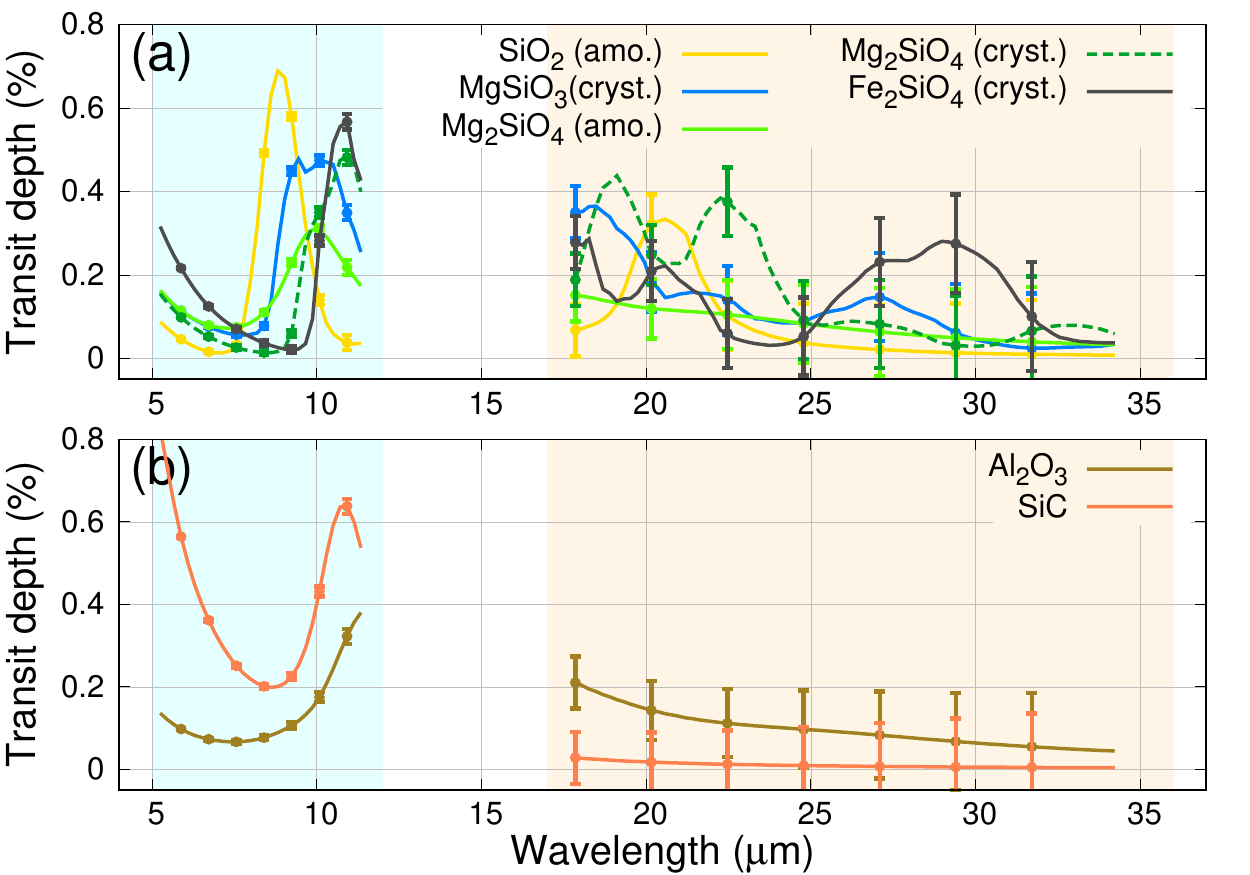}
\caption{{ Theoretical transmission spectra for candidate minerals shown in Figures~\ref{fig:th-spectra} (a) and (b)}, but here binned to $R = 10$. The error bars show expected $1\sigma$ noise for single transit observations with JWST MIRI/LRS (5--12 $\micron$; shaded in cyan) and SPICA SMI/LR (17--36 $\micron$; shaded in orange) assuming $d = 60~\rm pc$.
}
\label{fig:obs-spectra-Rp}
\end{figure}

For illustration, the error bars in Figure~\ref{fig:obs-spectra-Rp} indicate the $1\sigma$ observational uncertainties for the binned transmission spectra expected for $d=60$ pc.  
For this particular example, the signal level of the silicate features in the SPICA band is comparable to the noise level, suggesting that these features might be barely detectable. 
The detectability is more quantitatively discussed in the following subsections.

\subsection{Feature-to-noise Ratio} 
\begin{figure}[t]
\begin{center}
  \includegraphics[bb=0.000000 0.000000 355.000000 248.000000, width=8cm]{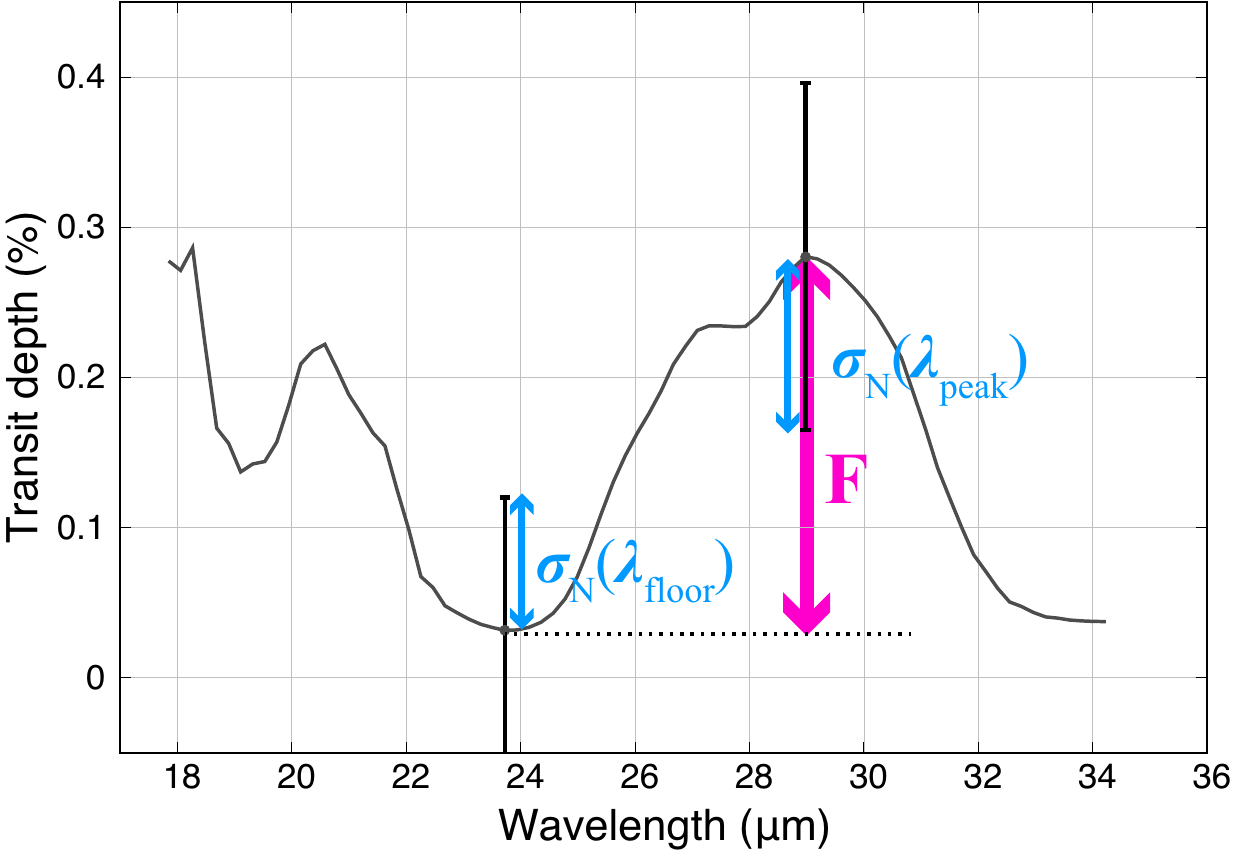}
\caption{Illustration showing the definition of ${\rm F} = \delta (\lambda_{\rm peak})- \delta (\lambda_{\rm floor})$,
$\sigma_{\rm N}(\lambda_{\rm peak})$, and $\sigma_{\rm N}(\lambda_{\rm floor})$ in the expression for ${\rm F/N}$ (Equation~\eqref{eq:FN}) for the case of the 24--32 $\micron$ peak of Fe$_2$SiO$_4$. The solid curve shows the binned theoretical transmission spectrum, while the vertical bars represent observational noise (see also Figure~\ref{fig:obs-spectra-Rp}).
\label{fig:FN-illust}}
\end{center}
\end{figure}

To quantify the significance of mineral features relative to observational noise, we introduce a metric, which we call the feature-to-noise ratio (F/N), defined as
\begin{equation}
{\rm F/N} = \frac{\delta (\lambda_{\rm peak})- \delta (\lambda_{\rm floor})}{\sqrt{{\sigma_{\rm N} (\lambda_{\rm peak})}^{2}+{\sigma_{\rm N} (\lambda_{\rm floor})}^{2}}}.
\label{eq:FN}
\end{equation}
where $\lambda_{\rm peak}$ is the wavelength at which the mineral feature under consideration is peaked and $\lambda_{\rm floor}$ is the wavelength at which the transit depth takes a local minimum just outside the feature.
In the right-hand side of Equation~\eqref{eq:FN}, the numerator is the relative transit depth of the feature, whereas the denominator is the root sum square of the  observational noises at $\lambda = \lambda_{\rm peak}$ and $\lambda_{\rm floor}$.
A feature with a higher F/N is more detectable.
For each mineral and for each of the two observational bands, we calculate the F/N of the strongest feature in the binned model spectrum.
Crystalline Mg$_2$SiO$_4$ and Fe$_2$SiO$_4$ are exceptional because they have multiple features that are comparable in magnitude. We take the 21--25 $\micron$ feature for the former and 24--32 $\micron$ feature for the latter.
Figure~\ref{fig:FN-illust} illustrates the definition of F/N works for the 24--32 $\micron$ feature of crystalline Fe$_2$SiO$_4$.

\begin{figure*}
    \centering
   \includegraphics[bb=0.000000 0.000000 894.134375 1069.069062, width=1.0\textwidth]{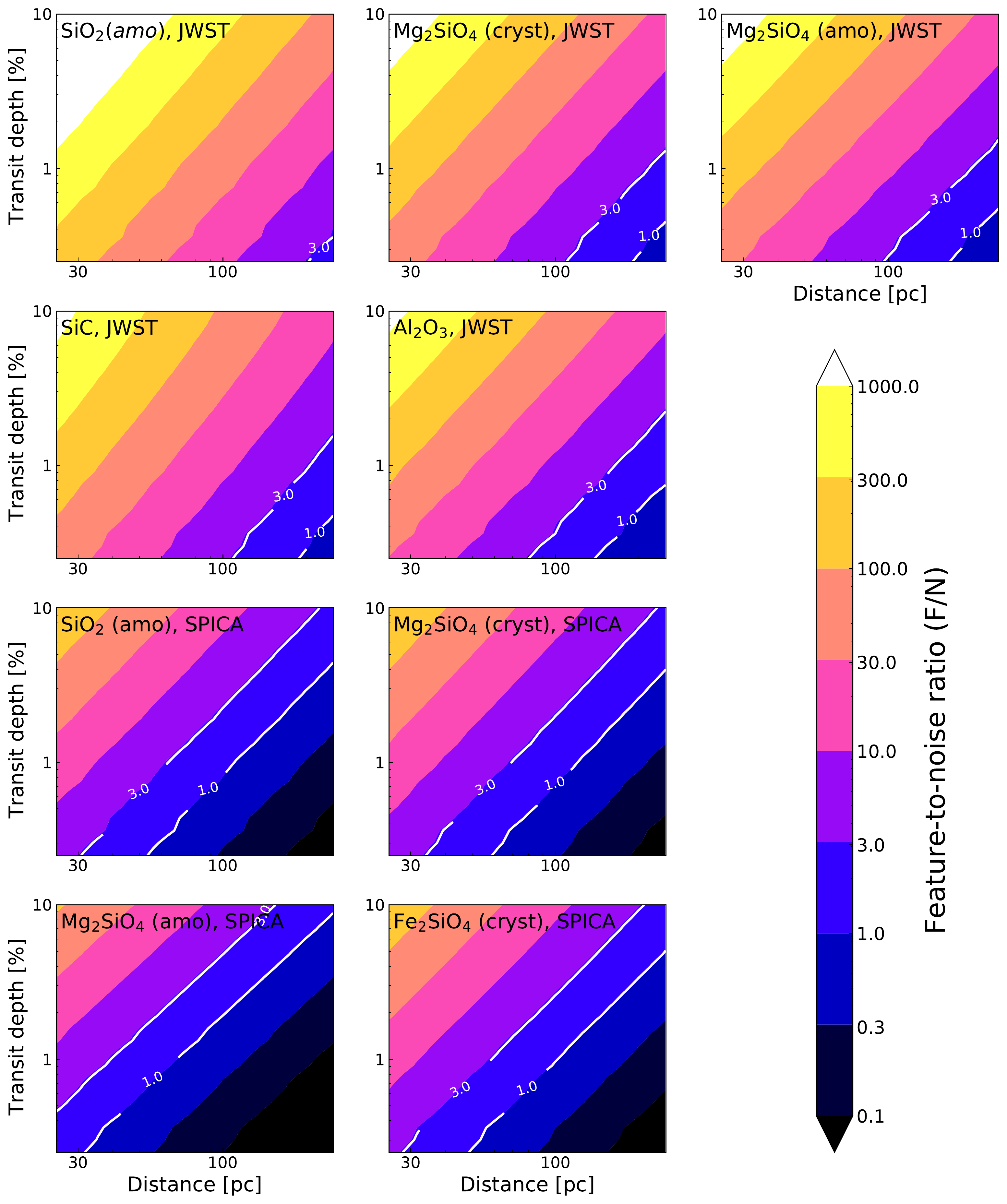}
    \caption{Feature-to-noise ratios (F/N) for different minerals from a single transit observation with JWST or SPICA as a function of the distance to the system, $d$, and the optical transit depth $\delta(\lambda_{\rm Kepler})$. The white contours mark ${\rm F/N} = 1$ and 3.
 }
 \label{fig:target-Rp}
\end{figure*}

\subsection{Detectability Prediction} 

We evaluate F/N for representative mineral features as a function of the transit depth in the Kepler band $\delta(\lambda_{\rm Kepler})$ and distance to the system $d$.
We vary $\delta(\lambda_{\rm Kepler})$ by adjusting the value of $R_{\rm t}$, while fixing $\alpha = 1/(0.133~\rm rad)$ and $n_0 = 5.6\times 10^4~\rm m^{-3}$ as in the previous section. 
As discussed in Section \ref{subsec:n}, adjusting the value of $n_0$ with fixed $R_{\rm t}$ gives essentially the same results for F/N, except that there is an upper limit on transit depth set by the tail's optical thickness.

Figure \ref{fig:target-Rp} shows the calculated F/N of each mineral for a single transit observation as a function of $\delta(\lambda_{\rm Kepler})$ and $d$.
Over the entire range of $\delta(\lambda_{\rm Kepler})$ explored here, the dust tail is optically thin both at optical and infrared wavelengths, and hence F/N increases monotonically with $\delta(\lambda_{\rm Kepler})$, or $R_{\rm t}$.
For planets at $d=100~{\rm pc}$,  the features of SiO$_2$, Mg$_2$SiO$_4$, SiC, and Al$_2$O$_3$ in the JWST band have F/N $\ga3$ if $\delta(\lambda_{\rm Kepler}) \ga 0.5\%$.
For those planets, the features of SiO$_2$, Mg$_2$SiO$_4$, and Fe$_2$SiO$_4$ in the SPICA band achieve F/N $\ga3$ if $\delta(\lambda_{\rm Kepler})$ exceeds $2$\%.

More specifically, taking $d=100~{\rm pc}$ and $\delta(\lambda_{\rm Kepler}) =1$\% as an example, one can detect the spectral features with F/N $>30$ for SiO$_2$, F/N $>10$ for Mg$_2$SiO$_4$ and SiC, and F/N $\sim$ 7 for Al$_2$O$_3$ in the JWST band. Among the mineral features in the spectra in the JWST band, 
SiO$_{2}$ has the largest feature at 8.8 $\mu$m, resulting in the highest F/N.
The absorption features of crystalline Mg$_2$SiO$_4$ and SiC are as detectable as amorphous Mg$_2$SiO$_4$.
Although the features of crystalline Mg$_2$SiO$_4$ and SiC, located at $\gtrsim 10\ \mu$m, are larger than those of amorphous Mg$_2$SiO$_4$, high observational noise of JWST at $\gtrsim 10\ \mu$m results in the similar ${\rm F/N}$ for these minerals.
The high observational noise yields the lowest ${\rm F/N}$ for the feature of Al$_{2}$O$_{3}$, located to $\sim12~{\rm \mu m}$, as compared to the other mineral features in the JWST band.

Overall, F/N for mineral features in the SPICA band tend to be lower than those for features in the JWST band. 
This is simply because the central star is fainter at longer wavelengths.
For the planet located at $d=100~{\rm pc}$ with $\delta(\lambda_{\rm Kepler}) =1$\%, one can observe the features with F/N $\sim$ 1 for SiO$_{2}$, crystalline Mg$_2$SiO$_4$, and crystalline Fe$_2$SiO$_4$, and F/N $\sim$ 0.3 for amorphous Mg$_2$SiO$_4$ in the SPICA band.
F/N $\sim$ 1 would be a minimum requirement to detect the features.
Therefore, one would need to observe the disintegrating planet with $\delta(\lambda_{\rm Kepler})\ga1$\% located at $d<100~{\rm pc}$ in order to detect the features in the SPICA band confidently.

\begin{figure*}
    \centering
    \includegraphics[bb=0.000000 0.000000 894.134375 1069.069062, width=1.0\textwidth]{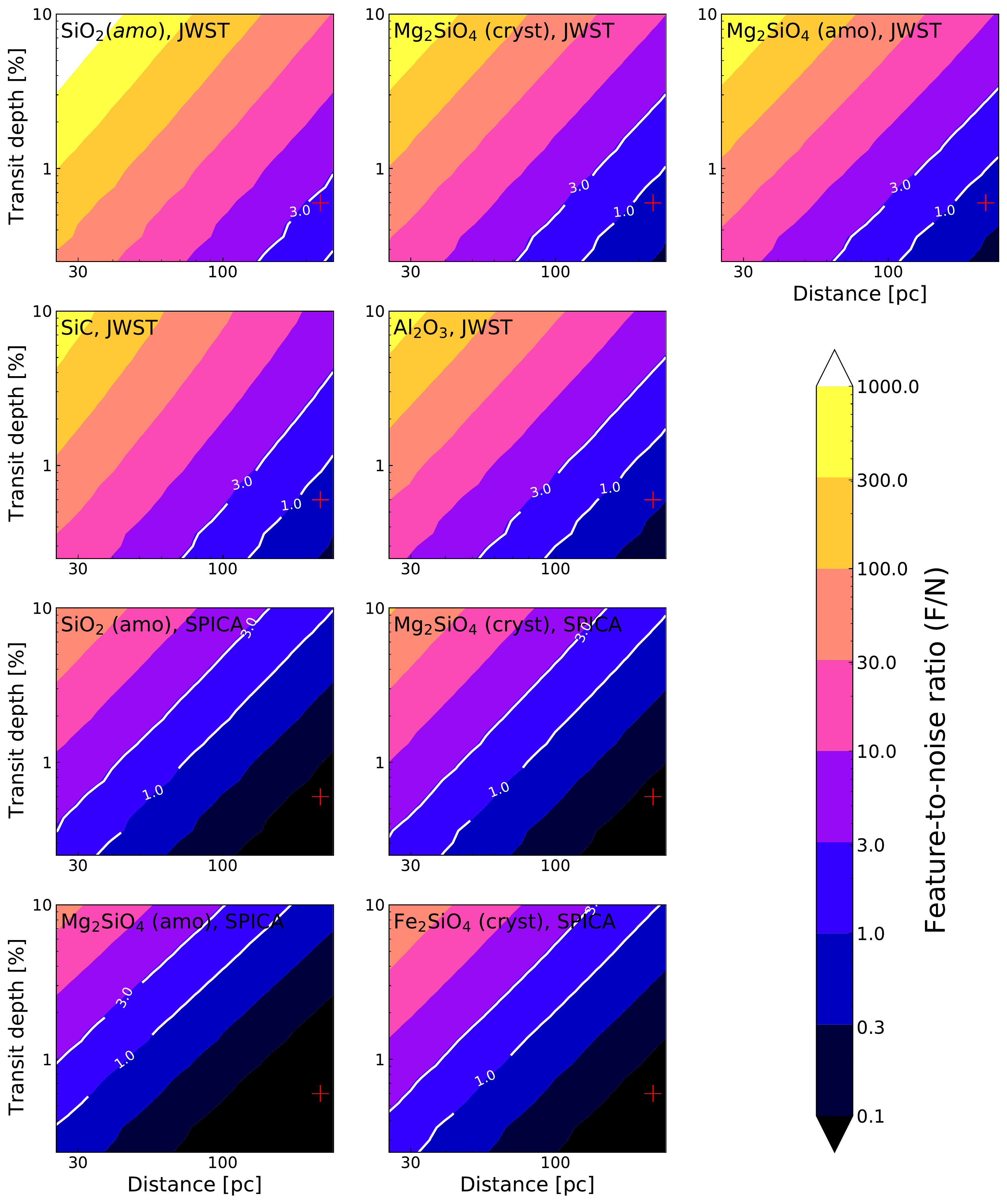}
    \caption{Same as Figure \ref{fig:target-Rp} except that the noise level is augmented by a factor of 5. The F/N shown here applies to targets lying near the ecliptic plane.
    K2-22b is placed on the red cross symbol.}
 \label{fig:target-Rp-high}
\end{figure*}

We here assess the detectability of mineral features for the nearest candidate of disintegrating exoplanets, K2-22b ($d$=225 pc and $\delta_{\rm Kepler} \sim$ 0.55\%).
Since K2-22b is located near the ecliptic plane, we adopt noise levels 5 times higher than for planets near the ecliptic pole (see Section~\ref{subsec:noise}). 
The F/N maps for this high noise case is shown in Figure \ref{fig:target-Rp-high}.
In a single transit observation of K2-22b with JWST, the feature of SiO$_2$ may be detectable with ${\rm F/N}\sim3$.
The other features in the JWST band have ${\rm F/N}\sim 1$, so detection of these features may  benefit from multiple transit observations.
In contrast, all the features in the SPICA band have F/N $< 0.1$, suggesting that detection of these features with SPICA is highly challenging for K2-22b.

Current and upcoming dedicated missions for transit searches such as {\it TESS} and PLATO would be able to provide optimal targets for the JWST and the SPICA transmission spectroscopy. Since N is smaller for closer planets, even small F and hence even small $\delta(\lambda_{\rm Kepler})$ provide the detection with F/N $\ga 3$. According to the estimation of \citet{Barclay+2018}, about 70\,\% of exoplanets found by {\it TESS} are expected to be located within as near as 200 pc.  
Moreover, {\it TESS} and PLATO will survey nearly the entire sky, allowing us to detect disintegrating planets far away from the ecliptic plane with high background noise. 
Although the optical transit depth is much smaller than that of K2-22b, a potential new candidate of the disintegrating planet has been recently discovered by {\it TESS} in the nearby system of DMPP-1 at 60 pc from the Earth \citep{Jones+2020}.

\section{Discussion} \label{sec:discussion}

\subsection{Effects of Varying $n_0$ on F/N}
\label{subsec:n}
In our model, we have produced dust tails of different optical transit depths $\delta(\lambda_{\rm Kepler})$  by varying the tail radius $R_{\rm t}$ while fixing the grain number density $n_{\rm 0}$ to be $5.6\times 10^4~\rm m^{-3}$.
Although this value of $n_{\rm 0}$ is consistent with the dust tail of K2-22b inferred to be optically thin (see Section~\ref{subsec:transit}), the choice is rather arbitrary.
For completeness, we study below how the countour map of F/N as shown in the previous section depends on the choice of $n_0$.

\begin{figure}[t]
   \includegraphics[bb=0.000000 0.000000 498.405000 340.868750, scale=0.5]{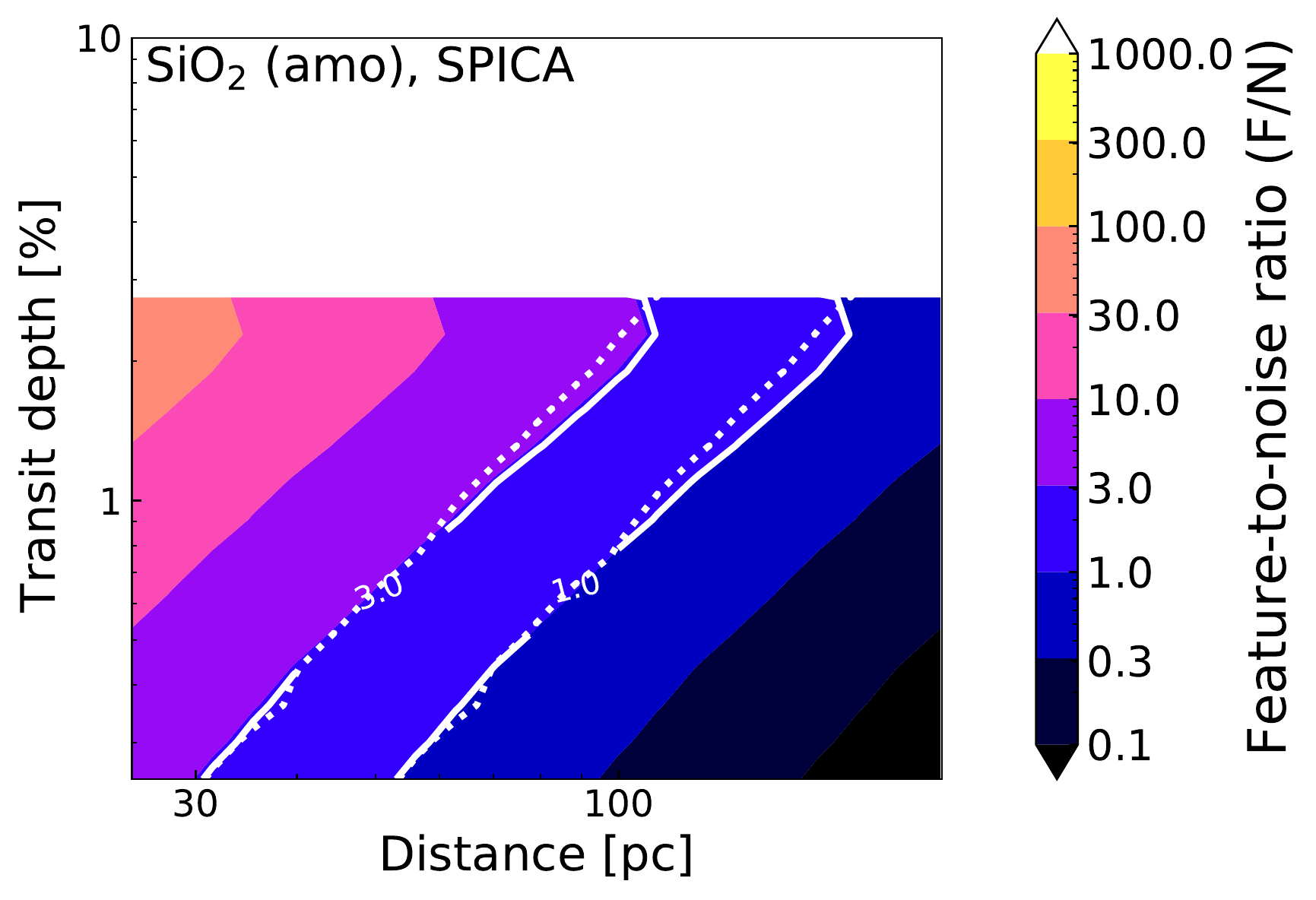}
    \caption{F/N map for amorphous SiO$_2$ obtained by varying $n_{0}$ to adjust the optical transit depth $\delta(\lambda_{\rm Kepler})$. The solid white lines mark F/N $=1$ and 3. For comparison, the white dotted line shows the contours of F/N $=1$ and 3 shown in the corresponding map in Figure~\ref{fig:target-Rp} (left column, third row).
 }
    \label{fig:target-n}
\end{figure}

We here take the feature of SiO$_{2}$ in the SPICA band as an example and calculate its detectability (F/N) by taking $n_{\rm 0}$ as a free parameter while fixing the tail radius to $R_{\rm t}=0.02R_{\rm star}$.
Figure \ref{fig:target-n} maps the values of F/N calculated for different sets of $n_0$ and $d$ on the $d$-$\delta(\lambda_{\rm Kepler})$ plane.
The map is bounded at  $\delta(\lambda_{\rm Kepler}) \approx 2.5\%$ because a dust tail of fixed $R_{\rm t}$ cannot have an arbitrary large projected area by varying $n_0$. 
In general, $\delta(\lambda_{\rm Kepler})$ increases linearly with $n_{\rm 0}$ as long as the tail is optically thin in the visible. However, as $n_{\rm 0}$ becomes so high that the tail becomes partially optically thick, $\delta(\lambda_{\rm Kepler})$ starts to saturate. 
In the limit of high optical thicknesses, the absorption cross section of the dust tail projected on the stellar surface approaches the geometric projected area $2R_{\rm t}\times 2R_{\rm star}$ (note that the tail length is assumed to be $\gg R_{\rm star}$; see Figure~\ref{fig:tail-model}), giving the upper limit on the optical transit depth of ($2R_{\rm t} \times 2R_{\rm star})/\pi {R_{\rm star}^{2}}$, which is $\approx 2.5\%$ for this example. 

Apart from the upper boundary, the map of F/N shown in Figure~\ref{fig:target-n} is approximately identical to that shown in Figure \ref{fig:target-Rp}, which was obtained by varying $R_{\rm t}$ while fixing $n_0$ (see the dashed lines in Figure~\ref{fig:target-n}).
This means that unless the tail is optically thick, the detectability of mineral features simply scales with the optical transit depth $\delta(\lambda_{\rm Kepler})$, and therefore either $R_{\rm t}$ or $n_0$ can be taken as a parameter controlling $\delta(\lambda_{\rm Kepler})$.

\subsection{Links to Planetary Composition} \label{subsec:links}
In this section, we discuss possible applications of the tail transmission spectroscopy to inferring the composition and formation process of disintegrating exoplanets.
We also review what formation processes are likely to yield these types of planets, and further discuss possible implications for planetary formation studies derived from their detection.
Based on the above discussion, we finally present how transmission spectra of the dust tails could distinguish between these types of planets with different formation background.

\subsubsection{Terrestrial Planets}  \label{subsec:terrestrial}
The crust, mantle, and core of differentiated rocky planets like Earth have distinct elementary compositions. 
The composition of the dust tail from such a planet would tell us about which part of the planet is currently evaporating.
For example, vapor of the bulk silicate Earth (BSE) elemental would form mineral clouds dominated by Mg$_2$SiO$_{4}$ and MgSiO$_{3}$ \citep{Mahapatra+2017}.
In contrast, vapor of a crust-like elemental composition would yield clouds of SiO$_{2}$ and SiO \citep[see Figure 11 of][]{Mahapatra+2017} because Si is an order of magnitude more abundant than Mg.
Vapor from crust-like materials may also yield dust dominated by Al, Ca, and Na-bearing minerals, which are abundant in the Earth's crust 
\citep{Rudnick+2003}.
A disintegrating planet that has already lost its crust and mantle would produce a tail of Fe metal grains.
We summarize the candidate composition of the dust tails produced by each part of a terrestrial planet in Figure \ref{fig:tail-comp}.

\subsubsection{Coreless Planets} \label{subsec:coreless}
Coreless terrestrial planets are hypothetical planets in which oxidization of iron inhibits the formation of an iron core  \citep{Elkins-Tanton+2008}.
In principle, such a planet could form if it initially contains some materials that can quickly oxidize iron before it sinks to to the planetary center.
Because such a planet likely contains Fe in addition to Si and Mg, one can expect that a disintegrating coreless planet forms a dust tail composed of iron-bearing silicates, such as Fe$_2$SiO$_{4}$, as well as Mg$_2$SiO$_{4}$ and MgSiO$_{3}$.
A transmission spectrum showing strong features of iron-bearing minerals will potentially serve as evidence for this (as yet hypothetical) type of planet.

Although how coreless planets can form is poorly understood, they could potentially form by accreting materials outside the H$_2$O snowline in protoplanetary disks.  
\citet{Elkins-Tanton+2008} suggested that a coreless planet could form when the planetary building blocks contain water-rich materials. 
Water oxidizes iron via metal-water reactions such as \citep[e.g.,][]{Lange+1984}
\begin{equation} 
{\rm Fe+ H_2O} \rightarrow {\rm FeO + H_2}.
\end{equation}
If such reactions proceed in the magma ocean rapidly, they could prevent the formation of an iron core \citep{Elkins-Tanton+2008}.  
Obviously, the building-block materials for a coreless planet would have to contain a significant amount of water. Therefore, if coreless disintegrating planets do exist, they would be most likely to have formed outside the snowline and then migrated to an ultra short-period orbit, or alternatively, formed in-situ by accreting oxidized materials delivered from outside the snowline.

 Depending on the initial water fraction and the oxidation rate of iron, some amount of water originally contained inside a protoplanet could be left intact. This is potentially important for observational identification of water-rich (but not necessarily coreless) disintegrating planets because the mantle of such planets may produce water vapor in addition to mineral vapor when it evaporates.
We come back to this point in Section~\ref{subsec:infer}.

\subsubsection{Carbon Planets \label{subsec:carbon}}
The mineralogy of  planets crucially depends on their C/O.
Carbon-rich (C/O $\ga1$) rocky planets could have a mantle of SiC and TiC and a crust of graphite  (e.g., \citealt{Kuchner&Seager2005,Bond+2010}; see Figure~\ref{fig:tail-comp}).
This is expected because if C/O $\ga1$, CO locks up the majority of the oxygen and thereby suppresses condensation of oxygen-bearing minerals \citep[e.g.,][]{Larimer75}.
The super-Earth 55 Cancri e is a candidate for such carbon planets \citep{Madhusudhan+2012,Willson&Militzer14}.
Carbon planets can potentially form in a carbon-rich disk around a carbon-rich star or in a particular disk location where carbon-rich grains are locally concentrated \citep[][see also references therein]{Kuchner&Seager2005}.
Silicate grains coated by soft organic mantles are sticky and may preferentially form carbon-rich planetesimals \citep{Kouchi+02,Homma+19}.

Because their formation environments require special conditions, carbon planets are commonly assumed to be rare.  
Data for white dwarf pollutions also suggest that the solids in extrasolar systems are depleted in carbon \citep[e.g.,][]{Gansicke+2012,Jura&Young14}.
Transmission spectroscopy of the dust tails of disintegrating planets will provide independent constrains on how rare carbon planets are. 
According to equilibrium mineral sequences for carbon-rich vapor \citep[e.g.,][]{Wood&hashimoto1993, Woitke+2018}, carbon disintegrating planets likely produce tails consisting mainly of SiC and graphite grains.

\subsubsection{Inferring the Planet Types from Tail Transmission Spectra \label{subsec:infer}}
Finally, we discuss how one can discriminate between terrestrial, coreless, and carbon disintegrating planets by looking at their transmission spectra.
If the silicates in their dust tails are crystalline,  combination of JWST and SPICA transmission spectra will allow us to infer their mineral composition (see also Figure~\ref{fig:obs-spectra-Rp}).
In the JWST band, silicates and SiC have an absorption feature at similar wavelengths $\sim10~{\rm \mu m}$.
In contrast, in the SPICA band, their spectral features are more diverse; Mg-bearing and Fe-bearing crystalline silicates showing features at $19$--$23~{\rm \mu m}$ and $26$--$30~{\rm \mu m}$, respectively, and  SiC is featureless.
Therefore, a tail transmission spectrum showing a feature at $\lambda\sim10~{\rm \mu m}$ in the JWST band and no feature in the SPICA band would be an indication of a carbon disintegrating planet.  
A tail spectrum exhibiting strong features at $\sim 10$ and $30~{\rm \mu m}$ may indicate a coreless planet containing abundant Fe in its mantle.

While our study has focused on dust, transmission spectroscopy of gas from disintegrating planets may also provide us with additional constraints of the planets' bulk compositions. 
For example, SiO vapor produces molecular absorption features at $\approx 7$--$10~\micron$ and $\ga 30~\micron$ \citep{Ito+2015}.
Water vapor also produces absorption features in the infrared wavelength ranges covered by JWST and SPICA \citep{Deming+2009,Greene+2016,Madhusudhan2019}. These molecular features could be useful to discriminate between disintegrating terrestrial planets and waterworlds. 
Since molecular features can be narrow in wavelength, their identification may require high-resolution spectroscopy.
Outside the midinfrared range, 
Lyman-$\alpha$ lines in the ultraviolet 
\citep[e.g.,][]{Vidal-Madjar+2003, Lecavelier+2010, Ehrenreich+2015} and
the absorption feature of excited metastable helium at 1083 nm \citep{Seager+2000, Oklopcic&Hirata18, Spake+2018}
are potentially useful to study whether the disintegrating planets possess an escaping atmosphere of hydrogen and helium.
Detection of the line absorption features of sodium and calcium atoms would give us constraints on the gas loss and sublimation processes at work on disintegrating planets \citep{Ridden-Harper+2019}.

\section{Conclusion} \label{sec:conclusion}
We have investigated the feasibility of constraining the composition of the dust tail
with combined spectroscopy with JWST MIRI and SPICA SMI.
We have constructed a simple model of the dust tail that calculates transit light curves and transmission spectra of disintegrating planets.

We have found that the combination of JWST (5--12 $\mu$m) and SPICA (17--36 $\mu$m) wavelength regions can provide stronger constraints on the composition of disintegrating planets than JWST alone (Table \ref{table:group} and Figure \ref{fig:th-spectra}).
Although crystalline and amorphous silicates, SiC, and Al$_2$O$_3$ have a feature at similar wavelengths in the JWST range, crystalline silicates have  additional features in the SPICA range that are absent for SiC and Al$_2$O$_3$. 
Moreover, crystalline Mg$_2$SiO$_4$ and crystalline Fe$_2$SiO$_4$ have peaks at different wavelengths in the SPICA range. 
Thus, observations with SPICA in addition to JWST would greatly help to disentangle the degeneracy of the candidate minerals of disintegrating planets.

We have also discussed how to infer the composition and formation process of disintegrating planets from the tail transmission spectra. According to previous studies, Mg-bearing minerals, Fe-bearing minerals, and SiC are potentially representative condensates of mantels of  terrestrial planets, coreless planets, and carbon planets, respectively.
One would be able to distinguish between these planets from tail transmission spectra in the JWST and SPICA bands because different types of disintegrating planets produce dust tails of different spectral features.

Based on the expected observational noise of the low-resolution spectrographs of JWST MIRI and SPICA SMI, we have quantified the detectability of absorption features of the candidate minerals in the tail transmission spectra at $R=10$. 
For a disintegrating planet located within the distance of 100 pc from the Earth with an optical depth deeper than $\sim 0.5$\%, tail transmission spectra in the JWST band show significant absorption features of minerals with feature-to-noise ratio of $\ga 3$, in a single transit (Fig \ref{fig:target-Rp}).
On the other hand, one can easily identify the mineral features in a single transit observation with SPICA if a disintegrating planet is located within 100 pc with a transit depth deeper than 2\%.
Multiple transit observations K2-22b's dust tail with JWST will allow us to constrain the minerals composition of the tail, although detection of mineral features with SPICA would be challenging for this particular object. 
Current and upcoming dedicated missions for transit searches, such as {\it TESS} and PLATO, would be able to provide ideal targets for future transmission spectroscopy with JWST and SPICA.
In this study, we have evaluated the detectability assuming the dust tails composed of a single mineral species, as a first step to examine the feasibility of such spectroscopic observations. The detectability in the case of the dust tails composed of multiple mineral species is left for future work.

\acknowledgments
We thank Akemi Tamanai for providing optical constant data, and Aki Takigawa, Shota Notsu, and Shigeru Ida for useful discussions. 
This work was supported by JSPS KAKENHI Grant Numbers JP18J14557, JP19K14783, JP18H05438, and JP19K03926.

\software{LX-MIE \citep{Kitzmann+2018}}




\newpage

\begin{thebibliography}{}
\expandafter\ifx\csname natexlab\endcsname\relax\def\natexlab#1{#1}\fi
\providecommand{\url}[1]{\href{#1}{#1}}
\providecommand{\dodoi}[1]{doi:~\href{http://doi.org/#1}{\nolinkurl{#1}}}
\providecommand{\doeprint}[1]{\href{http://ascl.net/#1}{\nolinkurl{http://ascl.net/#1}}}
\providecommand{\doarXiv}[1]{\href{https://arxiv.org/abs/#1}{\nolinkurl{https://arxiv.org/abs/#1}}}

\bibitem[{{Barclay} {et~al.}(2018){Barclay}, {Pepper}, \&
  {Quintana}}]{Barclay+2018}
{Barclay}, T., {Pepper}, J., \& {Quintana}, E.~V. 2018, \apjs, 239, 2,
  \dodoi{10.3847/1538-4365/aae3e9}

\bibitem[{{Bodman} {et~al.}(2018){Bodman}, {Wright}, {Desch}, \&
  {Lisse}}]{Bodman+2018}
{Bodman}, E. H.~L., {Wright}, J.~T., {Desch}, S.~J., \& {Lisse}, C.~M. 2018,
  \aj, 156, 173, \dodoi{10.3847/1538-3881/aadc60}

\bibitem[{{Bohren} \& {Huffman}(1983)}]{Bohren&Huffman83}
{Bohren}, C.~F., \& {Huffman}, D.~R. 1983, {Absorption and scattering of light
  by small particles}

\bibitem[{{Bond} {et~al.}(2010){Bond}, {O'Brien}, \& {Lauretta}}]{Bond+2010}
{Bond}, J.~C., {O'Brien}, D.~P., \& {Lauretta}, D.~S. 2010, \apj, 715, 1050,
  \dodoi{10.1088/0004-637X/715/2/1050}

\bibitem[{{Budaj}(2013)}]{Budaj2013}
{Budaj}, J. 2013, \aap, 557, A72, \dodoi{10.1051/0004-6361/201220260}

\bibitem[{{Deming} {et~al.}(2009){Deming}, {Seager}, {Winn}, {Miller-Ricci},
  {Clampin}, {Lindler}, {Greene}, {Charbonneau}, {Laughlin}, {Ricker},
  {Latham}, \& {Ennico}}]{Deming+2009}
{Deming}, D., {Seager}, S., {Winn}, J., {et~al.} 2009, \pasp, 121, 952,
  \dodoi{10.1086/605913}

\bibitem[{{Dorn} {et~al.}(2015){Dorn}, {Khan}, {Heng}, {Connolly}, {Alibert},
  {Benz}, \& {Tackley}}]{Dorn+15}
{Dorn}, C., {Khan}, A., {Heng}, K., {et~al.} 2015, \aap, 577, A83,
  \dodoi{10.1051/0004-6361/201424915}

\bibitem[{{Ehrenreich} {et~al.}(2015){Ehrenreich}, {Bourrier}, {Wheatley},
  {Lecavelier des Etangs}, {H{\'e}brard}, {Udry}, {Bonfils}, {Delfosse},
  {D{\'e}sert}, {Sing}, \& {Vidal-Madjar}}]{Ehrenreich+2015}
{Ehrenreich}, D., {Bourrier}, V., {Wheatley}, P.~J., {et~al.} 2015, \nat, 522,
  459, \dodoi{10.1038/nature14501}

\bibitem[{{Elkins-Tanton} \& {Seager}(2008)}]{Elkins-Tanton+2008}
{Elkins-Tanton}, L.~T., \& {Seager}, S. 2008, \apj, 688, 628,
  \dodoi{10.1086/592316}

\bibitem[{{Fabian} {et~al.}(2001){Fabian}, {Henning}, {J{\"a}ger}, {Mutschke},
  {Dorschner}, \& {Wehrhan}}]{Fabian+2001}
{Fabian}, D., {Henning}, T., {J{\"a}ger}, C., {et~al.} 2001, \aap, 378, 228,
  \dodoi{10.1051/0004-6361:20011196}

\bibitem[{{Fortney} {et~al.}(2007){Fortney}, {Marley}, \&
  {Barnes}}]{Fortney+2007}
{Fortney}, J.~J., {Marley}, M.~S., \& {Barnes}, J.~W. 2007, \apj, 659, 1661,
  \dodoi{10.1086/512120}

\bibitem[{{G{\"a}nsicke} {et~al.}(2012){G{\"a}nsicke}, {Koester}, {Farihi},
  {Girven}, {Parsons}, \& {Breedt}}]{Gansicke+2012}
{G{\"a}nsicke}, B.~T., {Koester}, D., {Farihi}, J., {et~al.} 2012, \mnras, 424,
  333, \dodoi{10.1111/j.1365-2966.2012.21201.x}

\bibitem[{{Glasse} {et~al.}(2015){Glasse}, {Rieke}, {Bauwens},
  {Garc{\'\i}a-Mar{\'\i}n}, {Ressler}, {Rost}, {Tikkanen}, {Vandenbussche}, \&
  {Wright}}]{Glasse+2015}
{Glasse}, A., {Rieke}, G.~H., {Bauwens}, E., {et~al.} 2015, \pasp, 127, 686,
  \dodoi{10.1086/682259}

\bibitem[{{Greene} {et~al.}(2016){Greene}, {Line}, {Montero}, {Fortney},
  {Lustig-Yaeger}, \& {Luther}}]{Greene+2016}
{Greene}, T.~P., {Line}, M.~R., {Montero}, C., {et~al.} 2016, \apj, 817, 17,
  \dodoi{10.3847/0004-637X/817/1/17}

\bibitem[{{Homma} {et~al.}(2019){Homma}, {Okuzumi}, {Nakamoto}, \&
  {Ueda}}]{Homma+19}
{Homma}, K.~A., {Okuzumi}, S., {Nakamoto}, T., \& {Ueda}, Y. 2019, \apj, 877,
  128, \dodoi{10.3847/1538-4357/ab1de0}

\bibitem[{{Ito} {et~al.}(2015){Ito}, {Ikoma}, {Kawahara}, {Nagahara},
  {Kawashima}, \& {Nakamoto}}]{Ito+2015}
{Ito}, Y., {Ikoma}, M., {Kawahara}, H., {et~al.} 2015, \apj, 801, 144,
  \dodoi{10.1088/0004-637X/801/2/144}

\bibitem[{{Jaeger} {et~al.}(1998){Jaeger}, {Molster}, {Dorschner}, {Henning},
  {Mutschke}, \& {Waters}}]{Jaeger+1998}
{Jaeger}, C., {Molster}, F.~J., {Dorschner}, J., {et~al.} 1998, \aap, 339, 904

\bibitem[{{Jones} {et~al.}(2020){Jones}, {Haswell}, {Barnes}, {Staab}, \&
  {Heller}}]{Jones+2020}
{Jones}, M.~H., {Haswell}, C.~A., {Barnes}, J.~R., {Staab}, D., \& {Heller}, R.
  2020, \apjl, 895, L17, \dodoi{10.3847/2041-8213/ab8f2b}

\bibitem[{{Jontof-Hutter}(2019)}]{Jontof-Hutter+2019}
{Jontof-Hutter}, D. 2019, Annual Review of Earth and Planetary Sciences, 47,
  141, \dodoi{10.1146/annurev-earth-053018-060352}

\bibitem[{{Jura} \& {Young}(2014)}]{Jura&Young14}
{Jura}, M., \& {Young}, E.~D. 2014, Annual Review of Earth and Planetary
  Sciences, 42, 45, \dodoi{10.1146/annurev-earth-060313-054740}

\bibitem[{{Kaneda} {et~al.}(2016){Kaneda}, {Ishihara}, {Oyabu}, {Yamagishi},
  {Wada}, {Kawada}, {Isobe}, {Asano}, {Suzuki}, {Nakagawa}, {Matsuhara},
  {Sakon}, {Tsumura}, {Shibai}, \& {Matsuo}}]{Kaneda+2016}
{Kaneda}, H., {Ishihara}, D., {Oyabu}, S., {et~al.} 2016, Society of
  Photo-Optical Instrumentation Engineers (SPIE) Conference Series, Vol. 9904,
  {SPICA Mid-infrared Instrument (SMI): technical concepts and scientific
  capabilities}, 99042I, \dodoi{10.1117/12.2232442}

\bibitem[{{Kitzmann} \& {Heng}(2018)}]{Kitzmann+2018}
{Kitzmann}, D., \& {Heng}, K. 2018, \mnras, 475, 94,
  \dodoi{10.1093/mnras/stx3141}

\bibitem[{{Kouchi} {et~al.}(2002){Kouchi}, {Kudo}, {Nakano}, {Arakawa},
  {Watanabe}, {Sirono}, {Higa}, \& {Maeno}}]{Kouchi+02}
{Kouchi}, A., {Kudo}, T., {Nakano}, H., {et~al.} 2002, \apjl, 566, L121,
  \dodoi{10.1086/339618}

\bibitem[{{Kuchner} \& {Seager}(2005)}]{Kuchner&Seager2005}
{Kuchner}, M.~J., \& {Seager}, S. 2005, arXiv e-prints, astro.
\newblock \doarXiv{astro-ph/0504214}

\bibitem[{{Lange} \& {Ahrens}(1984)}]{Lange+1984}
{Lange}, M.~A., \& {Ahrens}, T.~J. 1984, Earth and Planetary Science Letters,
  71, 111, \dodoi{10.1016/0012-821X(84)90057-8}

\bibitem[{{Larimer}(1975)}]{Larimer75}
{Larimer}, J.~W. 1975, \gca, 39, 389, \dodoi{10.1016/0016-7037(75)90204-5}

\bibitem[{{Lecavelier Des Etangs} {et~al.}(2010){Lecavelier Des Etangs},
  {Ehrenreich}, {Vidal-Madjar}, {Ballester}, {D{\'e}sert}, {Ferlet},
  {H{\'e}brard}, {Sing}, {Tchakoumegni}, \& {Udry}}]{Lecavelier+2010}
{Lecavelier Des Etangs}, A., {Ehrenreich}, D., {Vidal-Madjar}, A., {et~al.}
  2010, \aap, 514, A72, \dodoi{10.1051/0004-6361/200913347}

\bibitem[{{Lopez}(2017)}]{Lopez17}
{Lopez}, E.~D. 2017, \mnras, 472, 245, \dodoi{10.1093/mnras/stx1558}

\bibitem[{{Madhusudhan}(2019)}]{Madhusudhan2019}
{Madhusudhan}, N. 2019, \araa, 57, 617,
  \dodoi{10.1146/annurev-astro-081817-051846}

\bibitem[{{Madhusudhan} {et~al.}(2012){Madhusudhan}, {Lee}, \&
  {Mousis}}]{Madhusudhan+2012}
{Madhusudhan}, N., {Lee}, K. K.~M., \& {Mousis}, O. 2012, \apjl, 759, L40,
  \dodoi{10.1088/2041-8205/759/2/L40}

\bibitem[{{Mahapatra} {et~al.}(2017){Mahapatra}, {Helling}, \&
  {Miguel}}]{Mahapatra+2017}
{Mahapatra}, G., {Helling}, C., \& {Miguel}, Y. 2017, \mnras, 472, 447,
  \dodoi{10.1093/mnras/stx1666}

\bibitem[{{Oklop{\v{c}}i{\'c}} \& {Hirata}(2018)}]{Oklopcic&Hirata18}
{Oklop{\v{c}}i{\'c}}, A., \& {Hirata}, C.~M. 2018, \apjl, 855, L11,
  \dodoi{10.3847/2041-8213/aaada9}

\bibitem[{{Owen} \& {Adams}(2019)}]{Owen&Adams19}
{Owen}, J.~E., \& {Adams}, F.~C. 2019, \mnras, 490, 15,
  \dodoi{10.1093/mnras/stz2601}

\bibitem[{{Owen} \& {Wu}(2017)}]{Owen&Wu17}
{Owen}, J.~E., \& {Wu}, Y. 2017, \apj, 847, 29,
  \dodoi{10.3847/1538-4357/aa890a}

\bibitem[{{Perez-Becker} \& {Chiang}(2013)}]{Perez+2013}
{Perez-Becker}, D., \& {Chiang}, E. 2013, \mnras, 433, 2294,
  \dodoi{10.1093/mnras/stt895}

\bibitem[{{Rappaport} {et~al.}(2014){Rappaport}, {Barclay}, {DeVore}, {Rowe},
  {Sanchis-Ojeda}, \& {Still}}]{Rappaport+2014}
{Rappaport}, S., {Barclay}, T., {DeVore}, J., {et~al.} 2014, \apj, 784, 40,
  \dodoi{10.1088/0004-637X/784/1/40}

\bibitem[{{Rappaport} {et~al.}(2012){Rappaport}, {Levine}, {Chiang}, {El
  Mellah}, {Jenkins}, {Kalomeni}, {Kite}, {Kotson}, {Nelson},
  {Rousseau-Nepton}, \& {Tran}}]{Rappaport+2012}
{Rappaport}, S., {Levine}, A., {Chiang}, E., {et~al.} 2012, \apj, 752, 1,
  \dodoi{10.1088/0004-637X/752/1/1}

\bibitem[{{Ridden-Harper} {et~al.}(2019){Ridden-Harper}, {Snellen}, {Keller},
  \& {Molli{\`e}re}}]{Ridden-Harper+2019}
{Ridden-Harper}, A.~R., {Snellen}, I.~A.~G., {Keller}, C.~U., \&
  {Molli{\`e}re}, P. 2019, \aap, 628, A70, \dodoi{10.1051/0004-6361/201834433}

\bibitem[{{Roelfsema} {et~al.}(2018){Roelfsema}, {Shibai}, {Armus}, {Arrazola},
  {Audard}, {Audley}, {Bradford}, {Charles}, {Dieleman}, {Doi}, {Duband},
  {Eggens}, {Evers}, {Funaki}, {Gao}, {Giard}, {di Giorgio}, {Gonz{\'a}lez
  Fern{\'a}ndez}, {Griffin}, {Helmich}, {Hijmering}, {Huisman}, {Ishihara},
  {Isobe}, {Jackson}, {Jacobs}, {Jellema}, {Kamp}, {Kaneda}, {Kawada},
  {Kemper}, {Kerschbaum}, {Khosropanah}, {Kohno}, {Kooijman}, {Krause}, {van
  der Kuur}, {Kwon}, {Laauwen}, {de Lange}, {Larsson}, {van Loon}, {Madden},
  {Matsuhara}, {Najarro}, {Nakagawa}, {Naylor}, {Ogawa}, {Onaka}, {Oyabu},
  {Poglitsch}, {Reveret}, {Rodriguez}, {Spinoglio}, {Sakon}, {Sato},
  {Shinozaki}, {Shipman}, {Sugita}, {Suzuki}, {van der Tak}, {Torres Redondo},
  {Wada}, {Wang}, {Wafelbakker}, {van Weers}, {Withington}, {Vandenbussche},
  {Yamada}, \& {Yamamura}}]{2018PASA...35...30R}
{Roelfsema}, P.~R., {Shibai}, H., {Armus}, L., {et~al.} 2018, \pasa, 35, e030,
  \dodoi{10.1017/pasa.2018.15}

\bibitem[{{Rogers} \& {Seager}(2010)}]{Rogers&Seager10a}
{Rogers}, L.~A., \& {Seager}, S. 2010, \apj, 712, 974,
  \dodoi{10.1088/0004-637X/712/2/974}

\bibitem[{{Rudnick} \& {Gao}(2003)}]{Rudnick+2003}
{Rudnick}, R.~L., \& {Gao}, S. 2003, Treatise on Geochemistry, 3, 659,
  \dodoi{10.1016/B0-08-043751-6/03016-4}

\bibitem[{{Sakon} {et~al.}(2016){Sakon}, {Kaneda}, {Oyabu}, {Ishihara}, {Wada},
  \& {Fujishiro}}]{Sakon+2016}
{Sakon}, I., {Kaneda}, H., {Oyabu}, S., {et~al.} 2016, Society of Photo-Optical
  Instrumentation Engineers (SPIE) Conference Series, Vol. 9904, {Sensitivity
  estimates for the SPICA Mid-Infrared Instrument (SMI)}, 99043V,
  \dodoi{10.1117/12.2232402}

\bibitem[{{Sanchis-Ojeda} {et~al.}(2015){Sanchis-Ojeda}, {Rappaport},
  {Pall{\`e}}, {Delrez}, {DeVore}, {Gandolfi}, {Fukui}, {Ribas}, {Stassun},
  {Albrecht}, {Dai}, {Gaidos}, {Gillon}, {Hirano}, {Holman}, {Howard},
  {Isaacson}, {Jehin}, {Kuzuhara}, {Mann}, {Marcy}, {Miles-P{\'a}ez},
  {Monta{\~n}{\'e}s-Rodr{\'\i}guez}, {Murgas}, {Narita}, {Nowak}, {Onitsuka},
  {Paegert}, {Van Eylen}, {Winn}, \& {Yu}}]{Sanchis-Ojeda+2015}
{Sanchis-Ojeda}, R., {Rappaport}, S., {Pall{\`e}}, E., {et~al.} 2015, \apj,
  812, 112, \dodoi{10.1088/0004-637X/812/2/112}

\bibitem[{{Seager} {et~al.}(2007){Seager}, {Kuchner}, {Hier-Majumder}, \&
  {Militzer}}]{Seager+2007}
{Seager}, S., {Kuchner}, M., {Hier-Majumder}, C.~A., \& {Militzer}, B. 2007,
  \apj, 669, 1279, \dodoi{10.1086/521346}

\bibitem[{{Seager} \& {Sasselov}(2000)}]{Seager+2000}
{Seager}, S., \& {Sasselov}, D.~D. 2000, \apj, 537, 916, \dodoi{10.1086/309088}

\bibitem[{{Spake} {et~al.}(2018){Spake}, {Sing}, {Evans}, {Oklop{\v{c}}i{\'c}},
  {}, {Bourrier}, {Kreidberg}, {Rackham}, {Irwin}, {Ehrenreich}, {Wyttenbach},
  {Wakeford}, {Zhou}, {Chubb}, {Nikolov}, {Goyal}, {Henry}, {Williamson},
  {Blumenthal}, {Anderson}, {Hellier}, {Charbonneau}, {Udry}, \&
  {Madhusudhan}}]{Spake+2018}
{Spake}, J.~J., {Sing}, D.~K., {Evans}, T.~M., {et~al.} 2018, \nat, 557, 68,
  \dodoi{10.1038/s41586-018-0067-5}

\bibitem[{{van Lieshout} {et~al.}(2016){van Lieshout}, {Min}, {Dominik},
  {Brogi}, {de Graaff}, {Hekker}, {Kama}, {Keller}, {Ridden-Harper}, \& {van
  Werkhoven}}]{Lieshout+2016}
{van Lieshout}, R., {Min}, M., {Dominik}, C., {et~al.} 2016, \aap, 596, A32,
  \dodoi{10.1051/0004-6361/201629250}

\bibitem[{{Vidal-Madjar} {et~al.}(2003){Vidal-Madjar}, {Lecavelier des Etangs},
  {D{\'e}sert}, {Ballester}, {Ferlet}, {H{\'e}brard}, \&
  {Mayor}}]{Vidal-Madjar+2003}
{Vidal-Madjar}, A., {Lecavelier des Etangs}, A., {D{\'e}sert}, J.~M., {et~al.}
  2003, \nat, 422, 143, \dodoi{10.1038/nature01448}

\bibitem[{{Wilson} \& {Militzer}(2014)}]{Willson&Militzer14}
{Wilson}, H.~F., \& {Militzer}, B. 2014, \apj, 793, 34,
  \dodoi{10.1088/0004-637X/793/1/34}

\bibitem[{{Woitke} {et~al.}(2018){Woitke}, {Helling}, {Hunter}, {Millard},
  {Turner}, {Worters}, {Blecic}, \& {Stock}}]{Woitke+2018}
{Woitke}, P., {Helling}, C., {Hunter}, G.~H., {et~al.} 2018, \aap, 614, A1,
  \dodoi{10.1051/0004-6361/201732193}

\bibitem[{{Wood} \& {Hashimoto}(1993)}]{Wood&hashimoto1993}
{Wood}, J.~A., \& {Hashimoto}, A. 1993, \gca, 57, 2377,
  \dodoi{10.1016/0016-7037(93)90575-H}

\bibitem[{{Zeng} \& {Sasselov}(2013)}]{Zeng+2013}
{Zeng}, L., \& {Sasselov}, D. 2013, \pasp, 125, 227, \dodoi{10.1086/669163}

\bibitem[{{Zeng} {et~al.}(2019){Zeng}, {Jacobsen}, {Sasselov}, {Petaev},
  {Vanderburg}, {Lopez-Morales}, {Perez-Mercader}, {Mattsson}, {Li}, {Heising},
  {Bonomo}, {Damasso}, {Berger}, {Cao}, {Levi}, \& {Wordsworth}}]{Zeng+2019}
{Zeng}, L., {Jacobsen}, S.~B., {Sasselov}, D.~D., {et~al.} 2019, Proceedings of
  the National Academy of Science, 116, 9723, \dodoi{10.1073/pnas.1812905116}

\end{thebibliography}
\end{document}